\RequirePackage{amsthm}
\PassOptionsToPackage{margin=2cm}{geometry}
\documentclass[pdflatex,sn-mathphys-num]{sn-jnl}

\usepackage[table,xcdraw]{xcolor}
\usepackage{graphicx}%
\usepackage{multirow}%
\usepackage{amsmath,amssymb,amsfonts}%
\usepackage{amsthm}%
\usepackage{mathrsfs}%
\usepackage[title]{appendix}%
\usepackage{xcolor}%
\usepackage{textcomp}%
\usepackage{manyfoot}%
\usepackage{booktabs}%
\usepackage{algorithm}%
\usepackage{algorithmicx}%
\usepackage{algpseudocode}%
\usepackage{listings}%
\usepackage{booktabs} 
\usepackage{dcolumn}  
\usepackage{array}    
\usepackage{lineno}

\theoremstyle{thmstyleone}%
\theoremstyle{thmstyletwo}%

\theoremstyle{thmstylethree}%

\begin{document}

\title{Toxic behavior silences online political conversations}

\author*[1]{\fnm{Gabriela} \sur{Juncosa}}\email{juncosa\_maria@phd.ceu.edu}

\author[2,3]{\fnm{Taha} \sur{Yasseri}}\email{yasserit@tcd.ie}

\author[4,5]{\fnm{Júlia} \sur{Koltai}}\email{koltai.julia@tk.hun-ren.hu}

\author*[6,1,7]{\fnm{Gerardo} \sur{Iñiguez}}\email{gerardo.iniguez@tuni.fi}

\affil[1]{\orgdiv{Department of Network and Data Science}, \orgname{Central European University}, \orgaddress{\postcode{A-1100}, \city{Vienna}, \country{Austria}}}

\affil[2]{\orgdiv{School of Social Sciences and Philosophy}, \orgname{Trinity College Dublin}, \orgaddress{\postcode{D02 PN40}, \city{Dublin}, \country{Ireland}}}

\affil[3]{\orgdiv{Faculty of Arts and Humanities}, \orgname{Technological University Dublin}, \orgaddress{\postcode{D07 EWV4}, \city{Dublin}, \country{Ireland}}}

\affil[4]{\orgdiv{MTA–TK Lendület “Momentum” Digital Social Science Research Group for Social Stratification}, \orgname{HUN-REN Centre for Social Sciences}, \orgaddress{\postcode{1097}, \city{Budapest}, \country{Hungary}}}

\affil[5]{\orgdiv{Department of Social Research Methodology, Faculty of Social Sciences}, \orgname{ELTE Eötvös Loránd University}, \orgaddress{\postcode{1117}, \city{Budapest}, \country{Hungary}}}

\affil[6]{\orgdiv{Faculty of Information Technology and Communication Sciences}, \orgname{Tampere University}, \orgaddress{\postcode{FI-33720}, \city{Tampere}, \country{Finland}}}

\affil[7]{\orgdiv{Centro de Ciencias de la Complejidad}, \orgname{Universidad Nacional Autonóma de México}, \orgaddress{\postcode{04510}, \city{Ciudad de México}, \country{Mexico}}}

\abstract{Quantifying how individuals react to social influence is crucial for tackling collective political behavior online. While many studies of opinion in public forums focus on social feedback, they often overlook the potential for human interactions to result in self-censorship. Here, we investigate political deliberation in online spaces by exploring the hypothesis that individuals may refrain from expressing minority opinions publicly due to being exposed to toxic behavior. Analyzing conversations under YouTube videos from six prominent US news outlets around the 2020 US presidential elections, we observe patterns of self-censorship signaling the influence of peer toxicity on users' behavior. Using hidden Markov models, we identify a latent state consistent with toxicity-driven silence. Such state is characterized by reduced user activity and a higher likelihood of posting toxic content, indicating an environment where extreme and antisocial behaviors thrive. Our findings offer insights into the intricacies of online political deliberation and emphasize the importance of considering self-censorship dynamics to properly characterize ideological polarization in digital spheres. }


\maketitle
\newpage
\section{Introduction}\label{sec1}

As digital media become more widely used to engage with political content, \cite{gonzalez2023asymmetric,guess2023social}, online platforms increasingly cement their roles as key spaces for social interaction and the exchange of ideas. Digital media has fundamentally changed public communication by enabling individuals to create, share, and distribute content, as well as to engage more easily with others, promoting a level of connectivity that encourages deliberation \cite{gimmler2001deliberative, bressers2012mess}. It seems, however, that increased connectivity is a double-edged sword \cite{lorenz2023systematic}. While it pushes certain social movements forward, such as Black Lives Matter \cite{asmelash2020blm}, it also has the potential to fuel harmful activity, including the attack on the US Capitol in January 2021 \cite{nyt2021capitol, yasseri2023can}. This shift in connectivity transfers power to large digital media platforms, which now play a critical role in managing the digital infrastructure where these interactions occur. This influential position allows them to exert substantial control over both the infrastructure and the flow of information within it, raising key questions about data access and platform accountability \cite{stark2020algorithms, bak2021stewardship}.

Grasping the dynamics and quality of discussions shaped by the interconnected, algorithm-driven digital landscape is crucial in this context. Although the global shift to digital media has been associated with declining trust in politics \cite{porumbescu2017trust} and mainstream media \cite{guess2021partisan}, as well as with the rise of populism \cite{schumann2022spiral}, hate speech \cite{castano2021hate, vidgen2020detecting}, and increasing polarization \cite{bail2018polarization, yarchi2021digital}, it has also democratized access to information, enhanced political participation \cite{bond2012facebook, boulianne2020twenty, margetts2016political}, and has the potential to improve political knowledge \cite{beaudoin2008internet, park2018engament}. The existing literature presents conflicting views on the influence of digital media on political expression (see \cite{lorenz2023systematic} for a comprehensive review). However, this area of research is crucial, as evidence suggests that political expression on online platforms precedes political participation \cite{skoric2016expression}. As such, understanding how digital media may encourage or hinder diverse forms of communication in online forums is essential for leveraging its potential to promote productive discourse and diverse participation.

Although online communities in principle allow for constructive discussions in the face of political disagreement, the common occurrence of antisocial behavior—including insults, aggression, and ideological hostility—undermines their role in supporting tolerant deliberation in democratic societies \cite{dai2023addressing, oswald2023effects}. Harassment is a particularly large concern. A recent U.S. study shows that although the overall percentage of people experiencing online harassment remained stable at 41\% between 2017 and 2021, the incidence of severe harassment increased sharply from 15\% in 2014 to 25\% by 2021 \cite{vogels2022teens}. According to a related U.S. study, political views are the leading cause of online harassment, with one in five adults reporting harassment based on their political beliefs \cite{vogels2021online}. Toxic comments, including insults, are the most common forms of harassment encountered online \cite{thomas2021hate}. In the realm of online political discourse, toxicity not only amplifies extreme viewpoints and discourages moderate and marginalized voices, but it also deepens political divides, heightens polarization, and raises safety concerns that deter diverse participation in political discussions \cite{bail2022breaking, anderson2014nasty, kim2021distorting, bor2022psychology}. 

In this work, we use YouTube comments as a platform and the 2020 US presidential elections as a case study to explore the prevalence of toxic and insulting content in online political conversations. Even though toxicity is a widespread issue in online discussions across various platforms, its reported prevalence varies due to differing definitions and measurement approaches \cite{avalle2024persistent, kim2021distorting, oswald2023effects}. Here, we use Google’s Perspective API, a machine learning classifier trained on labeled data from sources like Wikipedia and the New York Times, to assess toxicity and insults \cite{wulczyn2017ex, lees2022new}. Our analysis focuses on two of the seven attributes of abusive comments that Perspective can identify: toxicity and insult. Perspective defines toxicity as a “rude, disrespectful, or unreasonable comment that is likely to make people leave a discussion”, and an insult as an “inflammatory, or negative comment towards a person or a group of people". The classifier assigns a probability score between 0 and 1 to indicate how likely a comment contains a given attribute \cite{perspective2024how}. While current automated toxicity detection systems face limitations mainly due to training dataset biases \cite{vidgen2020directions, medium2018bias, nogara2023toxic}, they remain an effective tool for conducting large-scale analyses at the level of entire populations \cite{avalle2024persistent, kim2021distorting}.

We aim to examine how toxic and insulting content emerges from user interactions and how this behavior impacts conversation dynamics and disengagement. We built a dataset of approximately 32.5 million comments and replies from videos posted by six prominent US news outlets on YouTube, chosen for their active comment sections and diverse political ideologies. Our dataset includes content from three left-leaning sources—ABC News, CBS, and CNN—and three right-leaning platforms: Newsmax, Fox News, and One America News Network. To focus on politically relevant discussions, we choose the time period from September 2020 to April 2021, covering key events surrounding the 2020 US presidential elections and the first 100 days of President Biden’s administration. Using a hidden Markov model (HMM) \cite{ramage2007hmm, eisner2002interactive, istrail2020hmm}, we study individuals' responses when confronted with a toxic and insulting environment. Our findings suggest that, in some contexts, toxic and insulting posts tend to surface toward the end of discussions. These findings provide insights into the complexities of online political deliberation and emphasize the importance of considering self-censorship dynamics in understanding digital discourse.

\section{Results}
\label{sec2}

\subsection {Toxic and insulting behavior appears across online political conversations}
\label{sec21}

We use YouTube as a case study to explore the prevalence of toxic and insulting content in online political discussions. Although YouTube is one of the largest and most engaging media platforms, it has received relatively limited academic attention compared to other social media \cite{hosseinmardi2021examining}. Previous research suggest YouTube’s role in fostering political radicalization and ideological echo chambers through personalized recommendations, finding that while content generally aligns with users' preferences, a notable right-wing bias leads far-right and moderate users to encounter more ideologically aligned and problematic content \cite{hosseinmardi2021examining, haroon2023auditing}. Other work has used YouTube data to identify consistent patterns of toxicity across different platforms and periods, finding that diverse opinions among users may contribute to rising toxicity levels \cite{avalle2024persistent}. Building on these efforts, our study provides an in-depth analysis of political conversations on YouTube, expanding the scope to include both toxic and insulting behaviors.

To focus on politically relevant discussions, we chose to study the 2020 U.S. presidential election. Our analysis period begins in September 2020—just over 60 days before Election Day—and extends through April 2021, covering Biden’s first 100 days in office. This timeframe captures a broad range of politically charged discussions both leading up to the election and during the early days of the new administration, a period traditionally marked by heightened media attention.

We selected the channels for our dataset by first analyzing media outlets' content and their engagement metrics, as listed in the 2019 AllSides Media Bias Chart \cite{mastrine2019bias}. We then narrowed our focus to six outlets with active comment sections that represent a broad spectrum of political ideologies: three left-leaning sources—ABC News, CBS, and CNN—and three right-leaning channels—Newsmax, Fox News, and One America News Network (OAN) (see \textbf{Appendix B} for information on bias ratings). Using the YouTube API, we collected top-level comments and replies on videos posted by these channels during our analysis period, along with engagement statistics for each video. The resulting dataset includes 18,627 videos and over 32 million comments and replies (see Table \ref{main:desciptive-stats}). 

We examine both individual comments and the conversations they form. A conversation is defined as a coherent, time-bound exchange of posts centered around a single top-level comment and its replies, capturing user interactions centered on a specific topic. Each conversation begins with a top-level comment posted on a YouTube video, which serves as the prompt for a conversation. All replies to this top-level comment are included in the conversation, ordered chronologically. To maintain relevance and focus, only replies posted within 10 days of the top-level comment are considered part of the conversation. YouTube supports only first-level replies, but users often work around this limitation by tagging or mentioning others in their comments to continue the conversation. It is possible to use these mentions to reconstruct sub-threads, but doing so is beyond the scope of this research. Using these mentions, it is therefore possible to reconstruct sub-threads; however, this is beyond the scope of this work. For the sake of simplicity, we assume that every reply targets the top-level comment, even if it mentions someone other than the comment's author. Consequently, each video contains multiple conversations, resulting in over 2.9 million conversations in our dataset. For detailed descriptive statistics, see Table \ref{main:desciptive-stats}.

To assess toxicity and insulting behavior, we use Google’s Perspective API, a machine learning classifier trained on labeled data from sources like Wikipedia and the New York Times \cite{wulczyn2017ex, lees2022new}. The Perspective API uses machine learning to detect abusive comments by scoring phrases based solely on their textual content, without considering emojis or images \cite{perspective2024how}. The model evaluates various attributes, including toxicity, insults, severe toxicity, identity attacks, threats, profanity, and sexually explicit content. However, we focus on toxicity and insults, as these attributes are the most prevalent in our dataset (see \textbf{Appendix C} for the prevalence of all attributes). Toxicity is defined as a “rude, disrespectful, or unreasonable comment that is likely to make people leave a discussion”, and insult as an “inflammatory, or negative comment towards a person or a group of people" \cite{perspective2024attributes}. 

The classifier assigns a probability score between 0 and 1 to indicate how likely a reader would perceive the comment as containing a given attribute. A higher score reflects a greater likelihood of the attribute being present. For example, a toxicity score close to 0 suggests the comment is unlikely to be toxic, a score of 0.5 indicates uncertainty , and a score of 1 means the comment is highly likely to be toxic \cite{perspective2024score}. Following previous research, we set our classification threshold at 0.6, and conducted robustness checks with different thresholds \cite{avalle2024persistent,saveski2021structure}.

While widely used, machine learning-based toxicity detection systems like the Perspective API have limitations due to their reliance on accurate and unbiased training datasets, which are challenging and resource-intensive to create \cite{vidgen2020directions}. Consequently, it is not surprising that the Perspective API may struggle to accurately identify and address toxicity in all contexts. For example, it has been found to assign higher toxicity scores to comments containing terms associated with frequently targeted groups (e.g., "Black," "Muslim," "feminist", "woman", "gay") \cite{medium2018bias}. Additionally, the classifier’s performance can vary by language; one study found that German content received significantly higher toxicity ratings, leading to nearly four times more moderated comments than in their English translations \cite{nogara2023toxic}. 

Despite these concerns, automated toxicity detection systems remain the most practical approach for large-scale analyses \cite{vidgen2020directions, sheth2022defining, avalle2024persistent}. Research has shown that the Perspective API performs comparably to human coders in classifying toxic comments on platforms such as Reddit \cite{rajadesingan2020quick}, and Facebook and Twitter \cite{hopp2020correlating, kim2021distorting}. Our examination of these examples suggests that, while not perfect, the Perspective API’s toxicity and insult scores generally correspond well with the content of the comments.

Table \ref{main:desciptive-stats} summarizes the key features of the dataset, with channels sorted according to their bias scores as determined by AllSides \cite{staff2023score}. The AllSides Media Bias Meter™ rates media outlets on a scale from -6 to +6, where 0 represents the Center, -6 indicates the farthest Left, and +6 denotes the farthest Right. This rating system replaces the previous five-category classification (Left, Lean Left, Center, Lean Right, and Right) to capture a more detailed view of media bias. Bias scores are determined using a combination of methods, including editorial reviews by a multipartisan panel and blind bias surveys, where participants rate outlets based on sampled content \cite{staff2023score}. We use the bias score to sort channels from left to right, an ordering that will be applied across all figures where applicable. 

We found that the overall prevalence of toxicity (10.1\%) and insulting language (10.4\%) in our dataset is relatively low. Yet, toxicity levels are higher than those reported in other YouTube studies, which report a prevalence of 4-7\% \cite{avalle2024persistent}. The higher prevalence of toxicity in our dataset may be attributed to the focus on politically charged topics, which tend to attract more emotionally charged discussions \cite{brady2017emotion}. Additionally, differences in sampling periods and the commenting culture of selected channels may contribute to these differences.

There is also significant variation across channels (Table \ref{main:desciptive-stats}). Left-leaning ABC News stands out with the highest level of toxicity (18.7\%) and insults (19.5\%), suggesting that its comment section may be among the most contentious. In contrast, CBS and right-leaning OAN show the lowest levels of toxic comments (5.6\% and 6.3\%, respectively) and insulting language (6.0\% and 5.9\%, respectively). A general trend emerges where right-leaning channels have lower average percentages of toxic and insulting comments compared to left-leaning channels. 

\begin{table}[t]
\caption{\textbf{Descriptive Statistics.} The dataset contains top-level comments and replies from videos posted by six prominent U.S. news outlets chosen for their active YouTube comment sections and for representing a broad spectrum of political ideologies. The time frame begins in September 2020 and extends through April 2021.}\label{main:desciptive-stats}
\begin{tabular}{@{}llcccccc@{}}
\toprule
                           & Channel  & \multicolumn{1}{l}{Bias Score\footnotemark[1]} & \multicolumn{1}{l}{Videos} & \multicolumn{1}{l}{Conversations} & \multicolumn{1}{l}{Comments} & \multicolumn{1}{l}{Toxicity\footnotemark[2]} & \multicolumn{1}{l}{Insult\footnotemark[3]} \\ \midrule
\multirow{3}{*}{Lean Left} & ABC News & -2.40                          & 4314                       & 453290                            & 4020626                      & 18.7                         & 19.5                       \\
                           & CBS      & -1.50                          & 4950                       & 361727                            & 2957079                      & 5.6                          & 6.0                        \\
                           & CNN      & -1.30                          & 1856                       & 895584                            & 10153723                     & 10.2                         & 9.5                        \\ \midrule
\multirow{3}{*}{Right}     & OAN      & 3.10                           & 1747                       & 49342                             & 613116                       & 6.3                          & 5.9                        \\
                           & Newsmax  & 3.28                           & 1752                       & 189679                            & 2844506                      & 7.0                          & 8.0                        \\
                           & Fox News & 3.88                           & 4008                       & 961907                            & 11890276                     & 9.2                          & 9.9                        \\ \midrule
                           & \multicolumn{2}{l}{ALL}                   & 18627                      & 2911529                           & 32479326                     & 10.1                         & 10.4                       \\ \bottomrule
\end{tabular}
\footnotetext[1]{
\href{https://www.allsides.com/blog/introducing-allsides-media-bias-meter}{https://www.allsides.com/media-bias/media-bias-chart}. Retrieved August 21, 2024}
\footnotetext[2]{Percentage of toxic comments. A comment is toxic if its toxicity score is greater than 0.6.}
\footnotetext[3]{Percentage of insulting comments. A comment is insulting if its insult score is greater than 0.6.}
\end{table}

\subsection {Spikes in activity, toxicity, and insults correlate with politically charged events} \label{sec22}

\begin{figure}[t]
\centering
\includegraphics[width=0.8\textwidth]{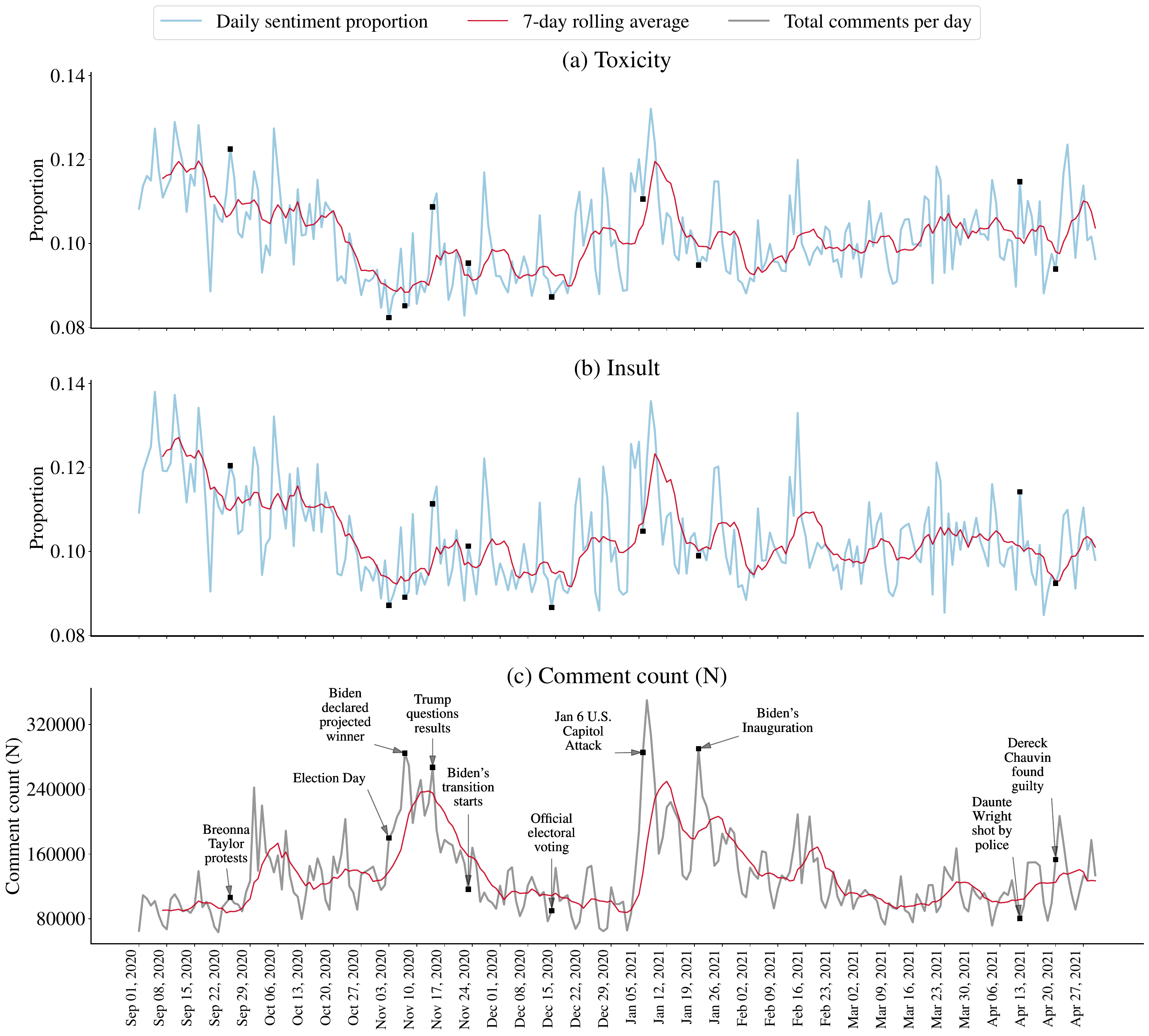}
\caption{\textbf{Trends in comment sentiment and volume over time.} The figure shows daily sentiment proportions (blue line), total number of comments per day (gray line), 7-day rolling average (red line) and relevant events annotated with black dots. \textbf{(a)} Proportion of toxic comments, defined as those with a toxicity score greater than 0.6. Peaks in toxicity are observed around September 2020, coinciding with heightened social and political tensions following the Black Lives Matter (BLM) protests. A subsequent peak occurs in January 2021, likely linked to the increased tensions surrounding the Capitol Attack. (b) Proportion of insulting comments, defined as those with an insult score greater than 0.6. The trend in insulting comments resembles that of toxicity; however, there is a more pronounced peak in February 2021, likely related to events surrounding the COVID-19 pandemic. (c) Total number of comments per day. A notable surge in activity coincides with the controversial events of the 2020 election; but, this increase is not mirrored in the trends for toxicity or insulting comments. The highest comment counts during the January 2021 Capitol Attack coincide with spikes in both toxicity and insults.}\label{timeseries}
\end{figure}

Moments of heightened national tension, such as controversial legal decisions, incidents of police violence, and election-related events, appear to trigger surges in negative online sentiment. Figure \ref{timeseries} shows time series data on the fraction of toxic and insulting posts (blue lines), daily comment counts (gray line), and a 7-day rolling average (red line). Key events are annotated with black dots. The proportion of toxic and insulting comments is defined as the fraction of comments with sentiment scores exceeding 0.6. Initially, both toxicity and insult trends exhibit relatively high averages, possibly reflecting social unrest and widespread demonstrations against racial injustice, particularly those connected to the Black Lives Matter movement. Notably, there is a peak in toxicity (Figure \ref{timeseries}, panel (a)) and a more modest increase in insults (panel (b)) and activity (panel (c)) following the grand jury’s decision to charge only one officer in the case of Breonna Taylor's death. This decision was followed by mostly peaceful nationwide demonstrations over the weekend of September 27–28, 2020 \cite{cnn2020btaylor}.

On April 11, 2021, Daunte Wright was fatally shot by police officer Kimberly Potter, an event that sparked protests across the U.S. and globally \cite{bbc2021blm}. These demonstrations overlapped with the ongoing investigation into George Floyd’s death at the hands of police officer Derek Chauvin. The timing of Wright’s death and Chauvin’s sentencing correlates with a rise in both toxicity and insults (panels (a) and (b)), alongside a modest increase in activity (panel (c)). 

Following Election Day, online activity continued to climb, with two notable peaks (panel (c)). The first spike coincided with Biden’s projected victory announcement, while the second occurred as Trump openly questioned the election's legitimacy. This latter moment also saw an increase in toxic and insulting posts (panels (a) and (b)). The dataset’s highest peak in activity appeared on January 6, 2021, the day of the Capitol Attack by Trump supporters, which marked the largest spike in toxicity and one of the highest in insulting posts. In February 2021, another distinct peak in insulting comments appeared, likely linked to events surrounding the COVID-19 pandemic, including vaccination rollout and debates over school reopening \cite{huang2021covid}. 

\subsection {Toxicity and insults fuel ongoing negativity} \label{sec23}
\begin{figure}[t]
\centering
\includegraphics[width=0.8\textwidth]{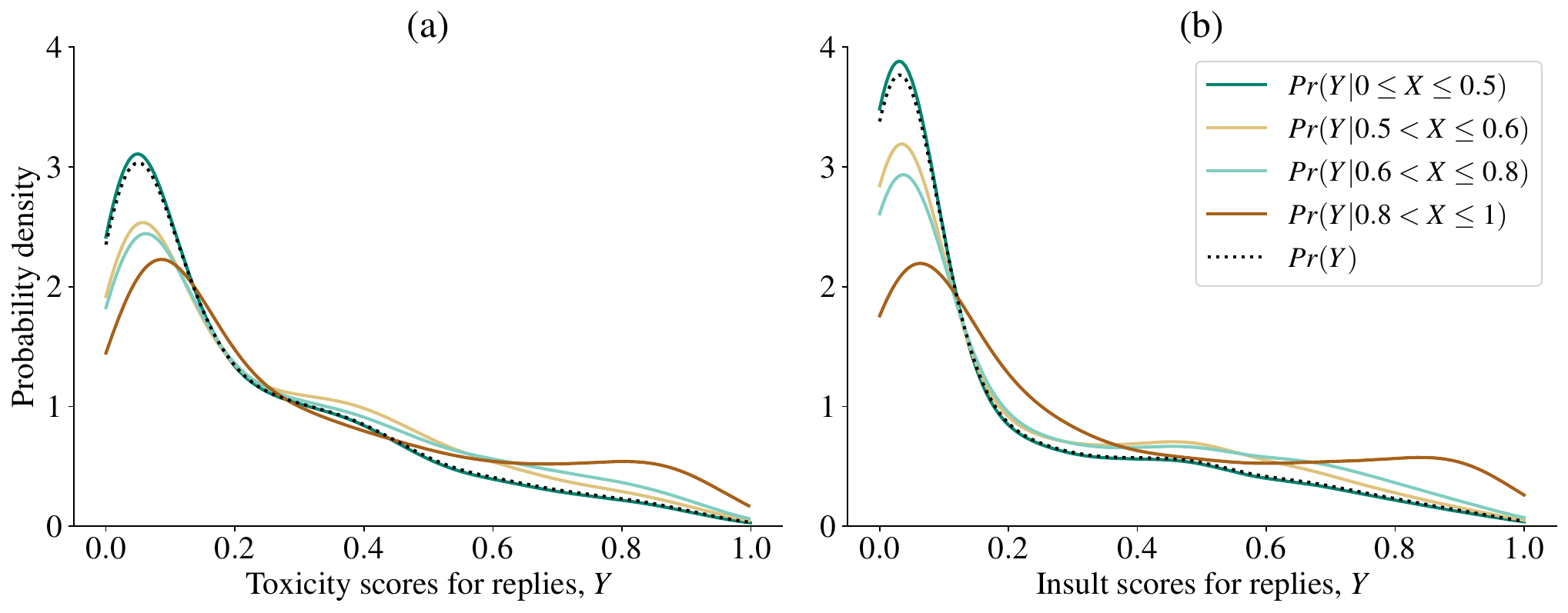}
\caption{\textbf{Probability density functions of sentiment scores for replies ($Y$) based on the sentiment scores of top-level comments ($X$).} The dotted line represents the unconditional distribution \( P(Y) \), showing the overall probability density of the sentiment scores for replies, while the solid lines indicate the conditional probabilities \( P(Y|X) \) for various ranges of $X$. Both panels show a correlation between the negative sentiment of top-level comments and their responses, with replies to highly toxic/insulting comments (brown lines) being more likely to have high toxicity/insult scores (greater than 0.8) than other replies \textbf{(a)} Probability density functions of toxicity scores. Replies are more likely to have toxicity scores between 0.2 and 0.6 than to have insult scores in the same range. \textbf{(b)} Probability density functions of insult scores. Replies are more likely to have low insult scores ($<$0.2) compared to low toxicity scores.}\label{marginalProbabilities}
\end{figure}

This section examines the relationship between sentiment scores for replies ($Y$) and the sentiment of corresponding top-level comments ($X$). Figure \ref{marginalProbabilities} illustrates the unconditional probability density function of sentiment scores for replies (dotted line) across two panels: the left panel (a) represents the probability density function of toxicity scores, while the right panel (b) shows the probability density function of insult scores. In both cases, the distributions are skewed to the left, peaking at low sentiment scores (below 0.2), suggesting that while both toxic and insulting top-level comments are likely to receive similarly negative responses, replies tend to have lower insult scores than toxicity scores. Additionally, replies are more likely to fall within the 0.2 to 0.6 range for toxicity scores than for insult scores. Finally, the areas under the unconditional probability density functions for scores exceeding 0.6 are similar for both toxicity and insult, aligning with prevalence estimates of 10.1\% for toxicity and 10.4\% for insult, as shown in Table \ref{main:desciptive-stats}.

In figure \ref{marginalProbabilities}, we also examine the probability density functions of sentiment scores for replies ($Y$) conditioned to  the sentiment scores of top-level comments ($X$). The solid lines show the conditional probabilities ($P(Y|X)$) across different ranges of ($X$). The shapes of the conditional probabilities indicate that while the sentiment scores for the replies correlate with the sentiment scores of the top-level comments they respond to, toxicity tends to have a wider spread. In contrast, insults cluster more tightly around lower values. 

In figure \ref{marginalProbabilities}, panel (a), the blue line represents top-level comments with low toxicity scores (0 to 0.5), showing that replies to less toxic comments tend to have a lower likelihood of high toxicity. In contrast, the yellow and light blue lines represent top-level comments with higher toxicity levels, indicating that as the toxicity of the initial comment rises, so does the probability of a toxic reply. This trend is further highlighted by the brown line, which shows that replies to highly toxic comments are more likely to reach high toxicity levels themselves, surpassing other conditional probabilities when $Y > 0.65$. This pattern holds for insulting scores as well, as shown in panel (b) of figure \ref{marginalProbabilities}. These findings hint at an influence of initial comment sentiment on the nature of replies, with elevated toxicity or insult scores in top-level comments associated with higher likelihoods of similarly negative replies. 

\subsection {Toxicity and insults end online conversations}
\label{sec24}

While analyzing individual comments can reveal insightful patterns, examining conversations as a whole is essential for understanding user interactions on YouTube. We aim to study the various stages a conversation goes through as it develops. For example, we are interested in whether toxicity encourages further toxicity or discourages participation. One approach to this question is to consider toxicity and insults as expressions of an unobserved, underlying conversational state. From this perspective, conversations can be seen as sequences of toxic and non-toxic, or insulting and non-insulting, exchanges unfolding over time. This sequential nature suggests that conversations may transition between different underlying states, which, though not directly observable, produce posts with distinct sentiment levels. To model these transitions and gain a deeper understanding of how toxicity or insults evolve within a conversation, we can use Hidden Markov Models (HMMs). HMMs allow us to infer the latent states driving these observable patterns (e.g., toxicity and insult), offering insights into how toxic or non-toxic exchanges progress and influence the flow of the conversation \cite{ramage2007hmm}.

Markov Models are an effective tool for handling time series data; however, they encounter limitations when analyzing sequences of states that are not directly observable. In such cases, only a probabilistic representation of these hidden states is available, limiting the model's ability to fully capture the underlying patterns \cite{ramage2007hmm}. The utility of Hidden Markov Models (HMMs) is best understood through examples. Consider a future climatologist studying historical global warming trends to reconstruct past weather patterns without direct records. Instead of weather data, the climatologist uncovers a diary detailing daily ice cream consumption. Recognizing the relationship between weather and ice cream consumption, the climatologist can employ HMMs to analyze this observable behavior and infer the underlying weather conditions that likely influenced it \cite{eisner2002interactive}. Similarly, in our study, we may lack direct information about the specific states a conversation undergoes. However, by observing external indicators—such as whether a comment is toxic or insulting—we can infer the latent states shaping the conversation.

Formally, a Hidden Markov Model consists of a sequence of observed variables, denoted as $X$, and a corresponding sequence of hidden states, denoted as $Z$. The observed variables $X$ represent data we can directly measure, such as the sentiment of a comment. The hidden states $Z$, however, are latent variables that represent the underlying conditions (e.g., conversational tone) influencing the observations. Each hidden state $Z$ generates an observation $X$ according to a probability distribution, and transitions between hidden states follow a Markov process, where the probability of transitioning to a new state depends solely on the current state. This framework allows us to model the progression of sentiment in conversations, capturing how hidden conversational states influence observable patterns over time \cite{ramage2007hmm}. 

\begin{figure}[t]
\centering
\includegraphics[scale=0.4]{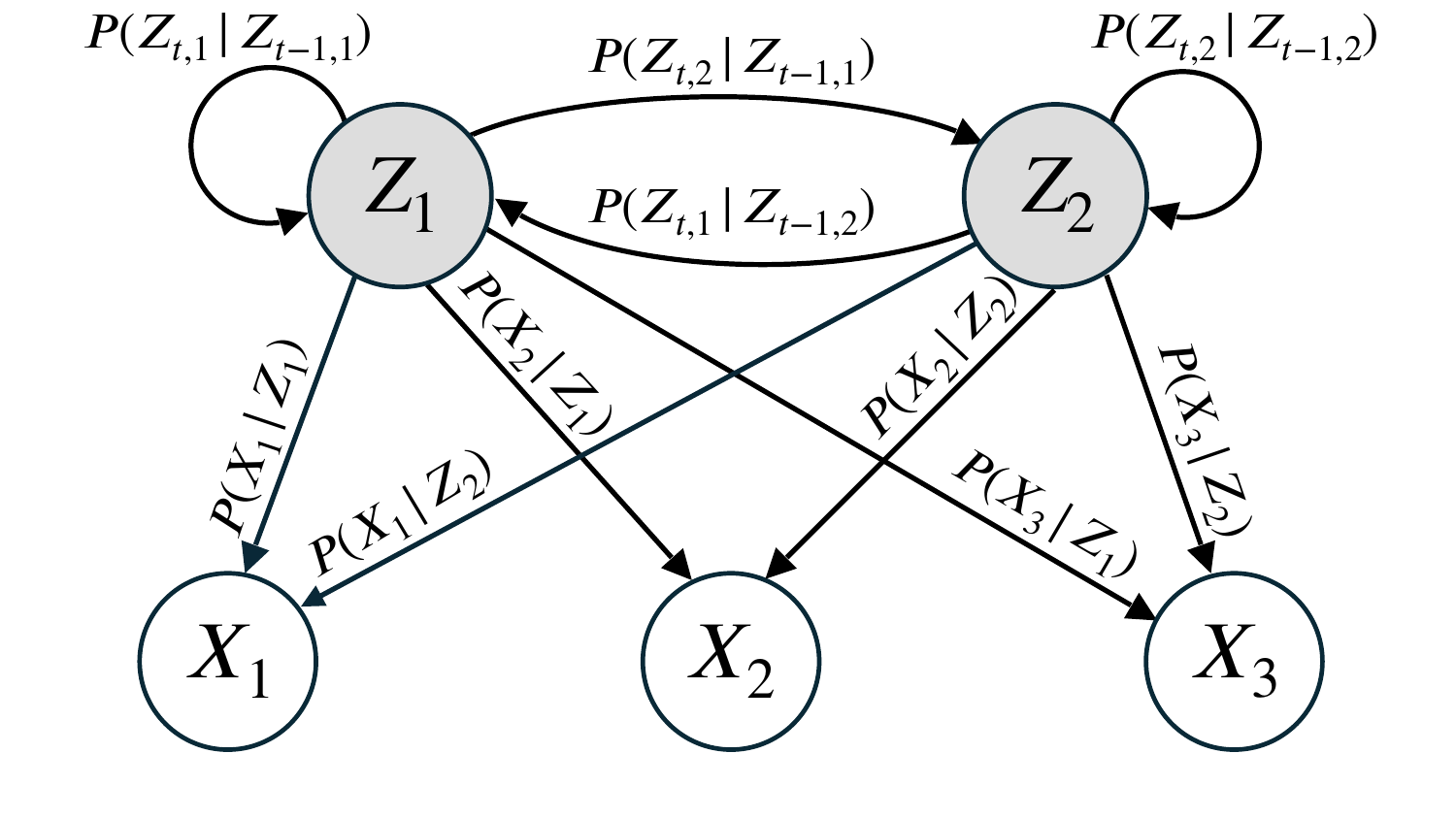}
\caption{\textbf{Diagram of the fitted Hidden Markov Model.} The model has two latent states (${Z}_{1}$ and ${Z}_{2}$) and three observed signals (${X}_{1}$, ${X}_{2}$ and ${X}_{3}$). The arrows represent the transition probabilities between latent states ($P({Z}_{t,i}|{Z}_{t-1,i})$ where $i=1,2$) and the emission probabilities of observed signals given the latent states ($P({X}_{j}|{Z}_{i})$ where $j=1,2,3$). The model captures the relationship between the unobserved states and the observed data over time.}\label{HMMDiagram}
\end{figure}

Figure \ref{HMMDiagram} illustrates an HMM with two latent states, ${Z}_{1}$ and ${Z}_{2}$, and three observed outcomes, ${X}_{1}$, ${X}_{2}$ and ${X}_{3}$. The top arrows depict the transition probabilities between latent states, $P({Z}_{t,i}|{Z}_{t-1,i})$ where $i=1,2$. These probabilities capture the likelihood of moving from one hidden state to another, defining the "rules" governing how states change over time. These probabilities are summarized in what is known as the transition matrix, an $I \times I$ row-stochastic matrix, where $P({Z}{t,1}|{Z}{t-1,1}) + P({Z}{t,2}|{Z}{t-1,1}) = 1$ and $P({Z}{t,2}|{Z}{t-1,2}) + P({Z}{t,1}|{Z}{t-1,2}) = 1$, ensuring that the total probability of transitioning from any state to another is always 1.

The bottom arrows in Figure \ref{HMMDiagram} represent the emission probabilities, $P({X}_{j}|{Z}_{i})$ where $j=1,2,3$, indicating the probability that a particular observation ${X}_{j}$ is generated given the system is in hidden state ${Z}_{i}$. Intuitively, each hidden state can be thought of as producing an observation with a certain probability. These probabilities are summarized in the emission matrix, an $I \times J$ row-stochastic matrix. This means that $P({X}_{1}|{Z}_{1})+P({X}_{2}|{Z}_{1})+P({X}_{3}|{Z}_{1})= 1$ and $P({X}_{1}|{Z}_{2})+P({X}_{2}|{Z}_{2})+P({X}_{3}|{Z}_{2})= 1$. 

HMMs can address three fundamental problems: evaluation, decoding, and learning. The \textit{evaluation} problem calculates the likelihood of a sequence of observed events given a particular HMM, useful in assessing how well the model fits a given observation sequence. The \textit{decoding} problem finds the most likely sequence of hidden states that could have generated a given sequence of observations, allowing us to infer the underlying state sequence responsible for the observed data. Finally, the \textit{learning} problem involves optimizing the model parameters—transition and emission probabilities—to best fit a set of observed sequences. This is typically achieved by iteratively adjusting these parameters to maximize the probability of the observations, thereby improving the model’s accuracy over time. These solutions make HMMs valuable for analyzing sequential data where the underlying states are hidden \cite{istrail2020hmm}.

In this work, we use HMMs to \textit{learn} the model parameters in different contexts. First, we learn model parameters for each of the six channels, that is, we infer the transition and emission probabilities that best align with the sequences of comments--conversations--in our dataset. Our modeling approach involves learning parameters for groups of videos, producing results that reflect aggregated behavior over video ensembles. In this section, we aggregate videos based on their news media channel publishers. We fitted the two-level HMM illustrated in Figure \ref{HMMDiagram}, where ${X}_{1}=0$ indicates lack of activity or the end of a conversation, ${X}_{2}=1$ represents a non-toxic or non-insulting post, and ${X}_{3}=2$ denotes a comment that is either toxic or insulting. Our primary focus is on the emission probabilities, as these will enable us to characterize and compare the hidden states. 

While we fixed the number of hidden states at two, we conducted robustness checks with models using three to five states (see SI section S4). These checks showed that the four-state model performed best. However, a qualitative comparison of the two-state and four-state models revealed that the four-state model primarily provides a more detailed characterization of state (${Z}{2}$), which represents the tone of the conversation preceding disengagement, where disengagement is not possible (i.e., $P({X}{1}|{Z}_{2}=0)$). While this refinement is informative, the primary aim of this study is to distinguish between self-censorship and active engagement. A detailed exploration of the variations within state ${Z}_{2}$ falls outside the scope of this work. We chose the simpler two-state model, guided by previous findings that HMM model selection methods frequently overestimate the optimal number of states \cite{pohle2017selecting}.

\begin{figure}[t]
\centering
\includegraphics[width=0.9\textwidth]{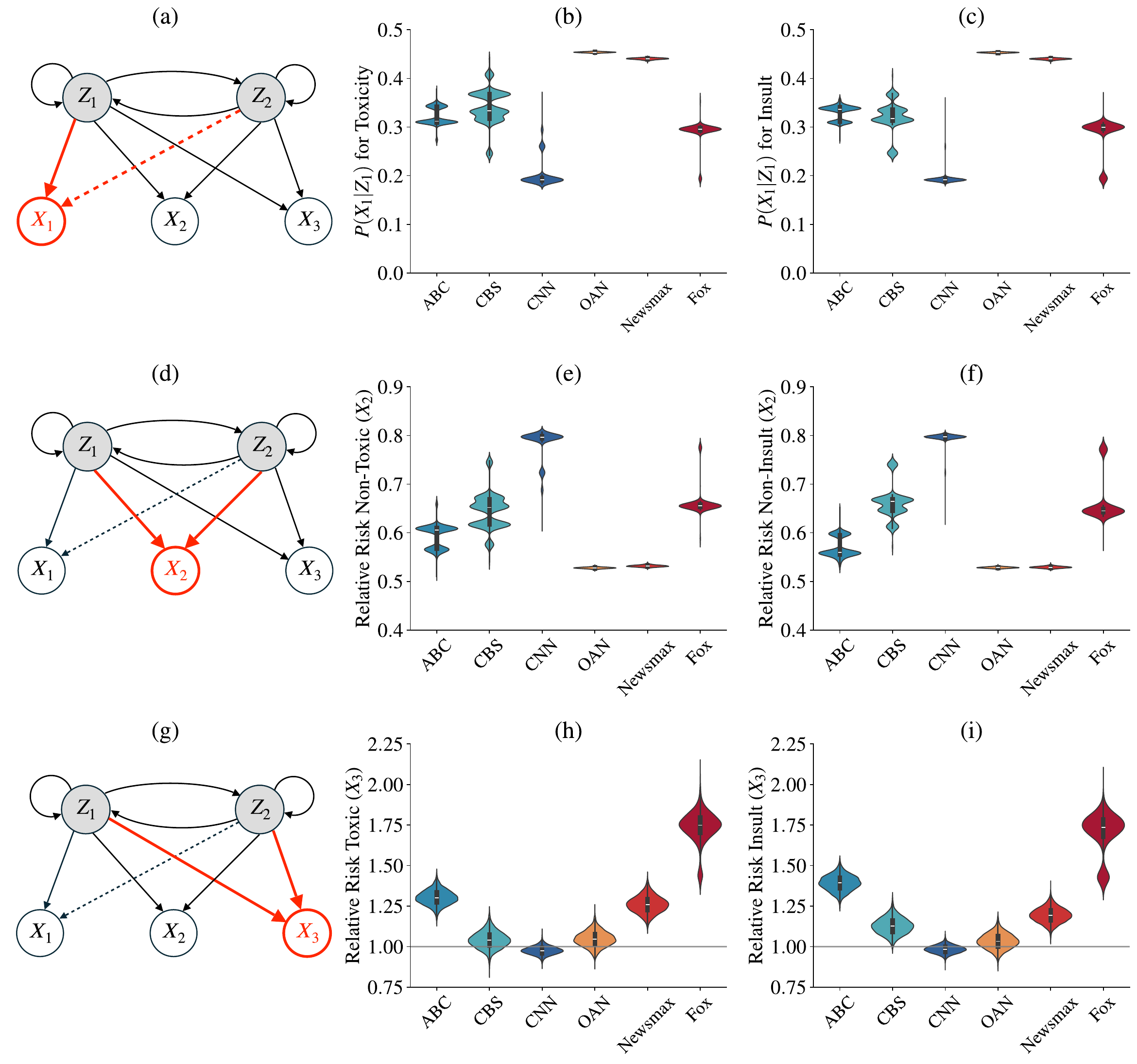}
\caption{\textbf{Inferred emission probabilities and relative risks by news channels.} Panels \textbf{(a)-(c)} show emission probabilities associated with ${X}{1}$, with panel \textbf{(b)} detailing $P({X}{1}|{Z}{1})$ for toxic content, panel \textbf{(c)} for insulting content, and panel \textbf{(a)} identifies the probabilities from Figure \ref{HMMDiagram} that are emphasized in panels (b) and (c). Notably, $P({X}{1}|{Z}{2}) = 0$ across all channels (indicated by a dotted arrow in panels \textbf{(a)}, \textbf{(d)}, and \textbf{(g)}), suggesting that conversations do not conclude in state ${Z}{2}$. This, combined with $P({X}{1}|{Z}{1}) > 0$, identifies state ${Z}{1}$ as the likely terminal state, while state ${Z}{2}$ corresponds to earlier stages in a conversation. Panels \textbf{(d)-(f)} display relative risk findings for ${X}{2}$, with non-toxic posts (panel \textbf{(e)}) and non-insulting posts (panel \textbf{(f)}). Here, $RR{{X}{2}} < 1$ across all channels, suggesting that non-toxic or non-insulting activity is more common in state ${Z}{2}$. Panels \textbf{(g)-(i)} summarize the relative risk findings for ${X}{3}$, with toxic posts in panel \textbf{(h)} and insulting posts in panel \textbf{(i)}. In almost all channels (with CNN as an exception), $RR{{X}{3}} > 1$, indicating that toxic or insulting posts are more likely when a conversation is in state ${Z}{1}$. This effect is most pronounced for Fox News (a right-leaning channel), while among left-leaning channels, ABC News exhibits the highest relative risk for ${X}_{3}$.}\label{HMMResultsChannel}
\end{figure}

After fitting the model for each channel, we found evidence of two distinct states through which conversations transition. Figure \ref{HMMResultsChannel} illustrates the inferred emission probabilities. Panels (a)-(c) present emission probabilities related to ${X}_{1}$: panel (b) displays $P({X}_{1}|{Z}_{1})$ for toxicity, panel (c) for insults across several channels, and panel (a) highlights the probabilities from Figure \ref{HMMDiagram} that are the focus of panels (b) and (c). Notably, $P({X}_{1}|{Z}_{2}) = 0$ for all channels, which is represented by a dotted arrow in the diagrams in panels (a), (d) and (g). Recall that ${X}_{1}$ denotes the end of a conversation; thus, $P({X}_{1}|{Z}_{2}) = 0$ indicates that conversations do not conclude in state ${Z}_{2}$. Combined with the fact that $P({X}_{1}|{Z}_{1}) > 0$ across all channels, we can interpret state ${Z}_{1}$ as the likely terminal state for conversations, while state ${Z}_{2}$ is associated with earlier stages of interaction.

To further characterize the difference between states ${Z}_{1}$ and ${Z}_{2}$, we analyzed the relative risk of non-toxic/non-insulting activity (${X}_{2}$) in state ${Z}_{1}$ compared to ${Z}_{2}$, denoted as ${RR}_{{X}_{2}}$, and the relative risk of toxic/insulting activity (${X}_{3}$) in state ${Z}_{1}$ versus ${Z}_{2}$, denoted as ${RR}_{{X}_{3}}$, where
\begin{equation}
    {RR}_{{X}_{2}} = \frac{P({X}_{2}|{Z}_{1})}{P({X}_{2}|{Z}_{2})}
\end{equation}
and, 
\begin{equation}
    {RR}_{{X}_{3}} = \frac{P({X}_{3}|{Z}_{1})}{P({X}_{3}|{Z}_{2})}
\end{equation}

A relative risk ${RR}_{{X}_{3}}>1$ suggests that negative sentiment is more likely to occur when a conversation is in state ${Z}_{1}$, whereas a relative risk ${RR}_{{X}_{3}}<1$ would indicate that negative behavior is more probable in state ${Z}_{2}$. Lastly, ${RR}_{{X}_{3}}=1$ means that negative sentiment is equally likely in both states. Similar interpretations apply to the relative risk of non-negative posts ${RR}_{{X}_{2}}$. 

Figure \ref{HMMResultsChannel}, panels (d)-(f), summarize the relative risk findings for ${X}_{2}$, specifically, non-toxic posts (panel (e)) and non-insulting posts (panel (f)). We observed that $RR_{{X}_{2}} < 1$ for all channels and both sentiment types, indicating that non-toxic or non-insulting activity is more likely to occur when a conversation is in the state ${Z}_{2}$—that is, when it is not nearing its end. 

Figure \ref{HMMResultsChannel}, panels (g)-(i) summarize results for relative risk ${X}_{3}$, where panel (h) shows results for toxic content across different news channels, organized from left-leaning to right-leaning, and panel (i) for insulting content. We found that $RR_{{X}_{3}} > 1$ for nearly all channels, with CNN as the exception, This pattern holds for both toxic and insulting activity. This finding indicates that when a conversation is in state ${Z}_{1}$, toxic and insulting content is more likely to appear—suggesting that conversations approaching their end are more likely to contain toxic or insulting posts. 

The channels exhibit a U-shaped pattern in relative risk values  ${X}_{3}$. Left-leaning channels ABC News and CBS have relative risk, ${RR}_{{X}_{3}}$, values slightly above 1, indicating a mild increase in the probability of toxic content when in state ${Z}_{1}$. However, the risk remains close to the baseline (1.0), suggesting that toxic content is only marginally more likely in terminal conversations. CNN displays a relative risk close to 1 or even slightly below, indicating that toxic content in terminal conversations is about as likely as in other stages, creating the 'bottom' of the U-shape.

Right-leaning channels—OAN, Newsmax, and Fox News—show an increasing trend in relative risk ${RR}_{{X}_{3}}$, with Fox News exhibiting a particularly high relative risk (around 1.75). Although Newsmax's risk level is comparable to that of ABC News, the overall trend suggests that the risk of toxic content when in state ${Z}_{1}$ increases as the bias rating moves further to the right, with Fox News showing the highest risk. The relative risk ${RR}_{{X}_{3}}$ for insulting content follows a similar pattern. The U-shape pattern, where toxicity and insult levels in the state ${Z}_{1}$ increase as bias ratings move away from the center, suggests a polarization effect. This indicates that both ends of the political spectrum exhibit relatively high levels of toxicity and insult in terminal conversations.

\subsection {Topic-Based clustering of video descriptions and their association with toxic and insulting content} \label{sec24}
By learning parameters for video groups, our model produces results that summarize the aggregated behavior of video ensembles. In the previous section, we grouped videos by their news media channel publishers and observed that negative sentiment tends is more likely to arise toward the end of conversations. To ensure that the observed connection between negative sentiment and disengagement is not dependent on group definitions, we propose an alternative way to define video ensembles. Here, we categorize videos by the topics they address, regardless of their channel sources. This approach leverages topic modeling, a scalable machine-learning technique that organizes and classifies text by identifying semantically related content \cite{vayansky2020review}.

We use a hierarchical stochastic block model (hSBM) \cite{gerlach2018network} approach to categorize videos by topics. Compared to LDA, a widely used standard in the field, hSBM offers a framework that leverages the similarities between topic modeling and community detection in complex networks and reinterprets topic modeling as a community detection problem by representing the word-document matrix as a bipartite network. By utilizing community detection techniques for topic modeling, the hSBM approach constructs a nonparametric Bayesian model grounded in a hierarchical stochastic block model (hSBM), successfully overcoming many of LDA's limitations \cite{gerlach2018network}.

\begin{figure}[t]
\centering
\includegraphics[width=\textwidth]{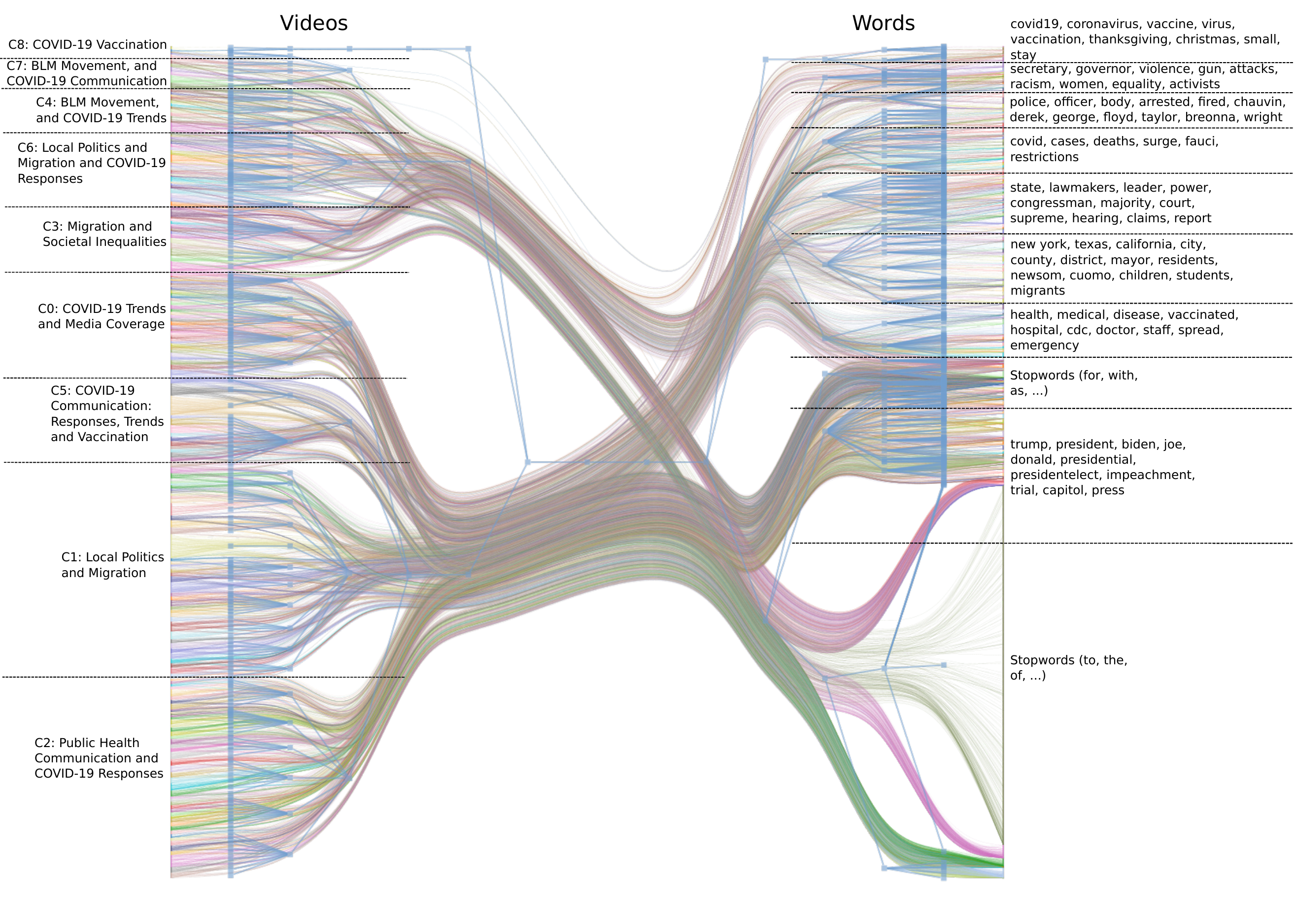}
\caption{\textbf{Inference of hSBM to video descriptions.} Results of clustering 19,365 video descriptions and 26,041 words, forming a network with 576,092 edges using the hSBM approach. At the highest hierarchical level, the model separates word nodes from video description nodes, reflecting its bipartite structure. At the fourth hierarchical level, it categorizes words into 10 topics, including two functional word topics (0 and 4), and identifies 9 distinct video clusters.}\label{main:youtube-VideoTopics}
\end{figure}

We use the hSBM approach (see Methods) to cluster 19,365 video descriptions and 26,041 words, creating a network comprising 576,092 edges. The model automatically identifies the number of topics and hierarchical levels, eliminating the need for prior specification. Figure \ref{main:youtube-VideoTopics} demonstrates the application of the hierarchical Stochastic Block Model (hSBM) on video descriptions. The clusters on the right side represent inferred topics, while the left side correspond to video groupings. At the highest level, the model separates word nodes from video description nodes, reflecting its bipartite structure. At the fourth level of hierarchy, it categorizes words into 10 topics, with two topics (0 and 4) representing functional words. (The model's ability to recognize function words provides a data-driven alternative to the traditional practice of manually removing stopwords.)

The remaining topics address various issues. Topics 1, 2, and 9 focus on matters related to the COVID-19 pandemic. Topics concerning U.S. national and local politics include the Supreme Court Nomination Controversy (topic 3), debates about migration (topic 5), the 2020 Presidential Election and the January 6 Capitol Attack (topic 6), and policy responses to societal inequalities (topic 8). Lastly, topic 7 pertains to issues of police brutality and the Black Lives Matter movement, a prominent subject at the time.

\begin{figure}[t]
\centering
\includegraphics[width=0.6\textwidth]{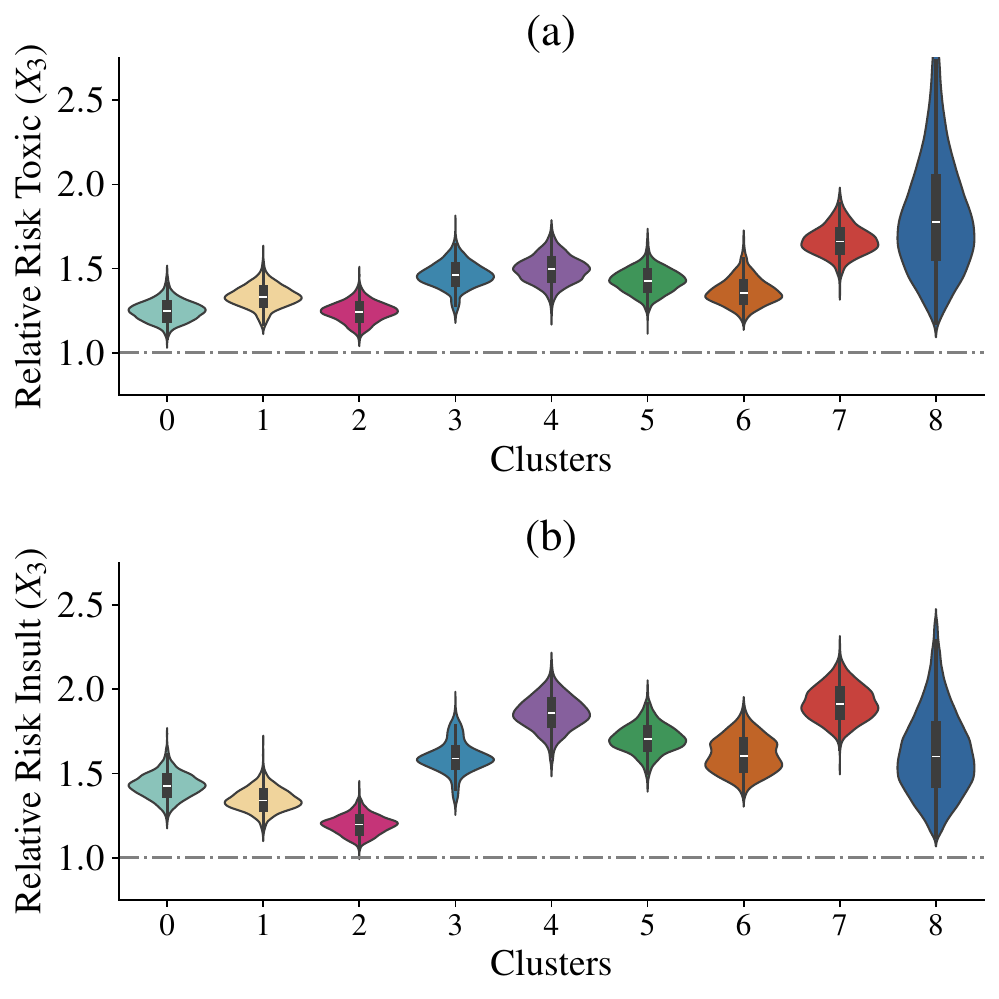}
\caption{\textbf{Inferred relative risks for toxic and insulting content by topic clusters.} \textbf{Clusters 1, 2, and 0:} Clusters focusing on the 2020 U.S. Presidential Election, January 6 Capitol Attack, and COVID-19 media coverage have low relative risks for toxic content (1.2–1.3) and insults, with cluster 0 slightly higher at just below ${RR}_{{X}_{3}}=1.5$. \textbf{Clusters 3 and 6:} Clusters discussing the Election and Capitol Attack, along with additional topics like migration, societal inequality, and COVID-19 communication, show low relative risks for toxic content but higher risks for insults when debates reach terminal stages, exceeding ${RR}_{{X}_{3}}=1.5$. \textbf{Clusters 5 and 8:} Clusters focusing on COVID-19 vaccination show contrasting behavior, where cluster 8 has the highest relative risk for toxic content, likely due to a small sample size with over representation of OAN videos, while cluster 5 has high insult risks exceeding ${RR}_{{X}_{3}}=1.5$, reflecting the controversial nature of vaccination discussions. \textbf{Clusters 4 and 7:} Videos addressing police brutality and the Black Lives Matter movement exhibit elevated risks for insults, with cluster 7 also showing heightened toxicity, underscoring the emotionally charged and contentious nature of these topics, particularly in the final stages of discussions.}\label{HMMResultsTopics}
\end{figure}

Unlike conventional models such as LDA, hSBM automatically clusters documents. This feature allows us to group video descriptions based on the topics they address. For this analysis, we work with the clusters at the fourth hierarchical level. Following the approach outlined in the previous section, we used these clusters to fit a Hidden Markov Model to each video cluster, as illustrated in Figure \ref{HMMDiagram}. Our primary focus is on the emission probabilities, which enable us to characterize and compare the hidden states effectively. Figure \ref{HMMResultsTopics} plots results for the relative risk for toxic and insulting content by topic cluster. Omitted emission probabilities reveal similar patterns as in Figure \ref{HMMResultsChannel}. 

Most of the video clusters focus on the 2020 U.S. Presidential Election and the January 6 Capitol Attack (topic 6), with clusters 1 and 2 being the most prominent in discussing this topic. Alongside these, cluster 0 addresses topics related to media coverage of COVID-19 pandemic trends and health outcomes. These three clusters exhibit similar and low relative risk for toxic content, with values ranging from 1.2 to 1.3. Similarly, these three clusters display comparable behavior in terms of the relative risk of insults, with the only distinction being that cluster 0 is slightly higher than the other two, reaching nearly  ${RR}_{{X}_{3}}=1.5$. 

Clusters 3 and 6 share similar profiles with clusters 1 and 2, as they also heavily discuss the Election and Capitol Attack. However, these clusters distinguish themselves by incorporating discussions on additional topics: migration debates and societal inequality issues (cluster 3), and local politics, migration, and COVID-19 communication (cluster 6). The relative risk of toxic content remains low for these clusters, with cluster 6 showing values comparable to cluster 1, while cluster 3 is slightly higher but still below ${RR}_{{X}_{3}}=1.5$. However, the relative risk of insults reaches similar levels for both clusters and exceeds ${RR}_{{X}_{3}}=1.5$, suggesting that videos addressing social debates, such as migration policies and societal inequality issues, are more likely to attract insulting posts as discussions approach their terminal stage.

Clusters 5 and 8 focus on topics related to the COVID-19 pandemic, particularly on vaccinations. Cluster 5 does not exhibit notable behavior regarding the relative risk of toxic content. Still, cluster 8 shows the highest relative risk value among all topics, although the variance is also significantly high. This is likely because cluster 8 is a small group with an overrepresentation of videos from OAN. While the relative risk of insults for cluster 8 does not maintain a high ranking, the values for cluster 5 exceed ${RR}_{{X}_{3}}=1.5$ and are among the highest. This suggests that content related to COVID-19 vaccination efforts tends to be more toxic (cluster 8) and more insulting (cluster 5) when conversations are nearing their end. This observation aligns with the intuition that COVID-19 vaccinations were a highly controversial topic that provoked significant contention.

Finally, clusters 4 and 7 touch upon issues of police brutality and the Black Lives Matter movement. It is worth noting that other than cluster 8, cluster 7 is the only other with a relative risk of toxic activity above 1.5, and it retains a high ranking when looking at the relative risk of insults $1.8<{RR}_{{X}_{3}}<2.0$. Likewise, cluster 4 displays comparable levels of relative risk of insults to cluster 7, although not for toxicity. These results suggest that videos discussing police brutality and the Black Lives Matter movement are more likely to host insulting content in the final stages of the conversation, highlighting the contentious and emotionally charged nature of these discussions.

All in all, by exploring an alternative categorization based on the topics addressed by videos, independent of their channel sources, we observe similar patterns across these different ensemble definitions, thereby confirming the robustness of the empirical relationship between negative sentiment and disengagement.

\section{Discussion}
\label{sec3}
Political segregation and the lack of conversation across ideological divides have become increasingly prevalent phenomena, posing significant risks to democratic discourse and the institutions underpinning it \cite{lorenz2023systematic, bak2021stewardship}. Social media platforms, as key players in modern communication, occupy a central role in this process \cite{blex2022positive}. These platforms, driven by algorithms designed to maximize engagement and monetary incentives, often create environments where political ideology clashes with marketing goals and user interactions \cite{stark2020algorithms}. A key outcome of such environments is the proliferation of antisocial behavior, such as toxicity and insults, primarily driven by small but vocal minorities. These behaviors significantly influence broader participation, fostering disengagement and self-censorship among the majority.

Our analysis of YouTube during the 2020 US presidential elections reveals several important findings. First, toxic and insulting behaviors are prevalent in online political conversations, particularly during politically charged events. We observed significant spikes in toxicity and insults coinciding with major political moments such as the Black Lives Matter protests, Election Day, and the Capitol riots of January 6. Furthermore, toxic top-level comments are more likely to elicit similarly negative replies, suggesting a cyclical relationship where toxicity fuels further toxicity. These patterns illustrate how toxicity is not an isolated phenomenon but an entrenched feature of political discourse online, particularly during moments of increased political tension. Together, these findings indicate that toxicity in online spaces reflects societal instability and is a driver of further polarization and disengagement.

The implications of these findings are far-reaching for understanding the dynamics of online political discourse. The spikes in toxicity observed during politically significant events suggest that online discourse is highly responsive to real-world political developments. This responsiveness reflects the emotional and ideological intensity of these moments, with online platforms serving as spaces for both expression and conflict \cite{brady2017emotion}. At the same time, the observation that toxicity generates further toxicity highlights how these platforms may reinforce and amplify negative interactions.

The cyclical nature of toxicity is particularly concerning because it can alienate a large portion of the user base. When individuals encounter toxic environments, they may disengage, self-censor, or even abandon the platform entirely. This dynamic reduces the diversity of perspectives in online conversations, skewing public discourse and misrepresenting public opinion. As such, online toxicity reflects polarization and actively contributes to it, undermining the potential of digital spaces to serve as arenas for democratic deliberation \cite{avalle2024persistent, falkenberg2024patterns}.

While this study provides meaningful insights, it is important to acknowledge its limitations. First, our analysis focuses on specific YouTube communities belonging to US news outlets, which, while providing detailed insights into conversations surrounding key political events in that country, narrows the scope of our results. The choice to study particular communities was intentional, allowing us to delve deeply into relevant discussions. However, the findings may not fully represent broader online discourse or apply to different contexts.

Second, although YouTube hosts a vast number of channels—many of which promote politically relevant content such as far-right ideologies, conspiracy theories, and "anti-woke" material—the scale and diversity of this content are beyond the scope of this project \cite{hosseinmardi2021examining}. Capturing a more comprehensive picture of all politically relevant conversations on YouTube would require datasets that include a broader range of independent creators and communities. This limitation suggests that future research should aim to expand datasets and examine a wider array of political content and interactions across different platforms.

Additionally, the influence of algorithmic recommendation systems and content moderation technologies adds complexity to the relationship between toxicity and self-censorship \cite{haroon2023auditing}. The role of these technologies in shaping user behavior and interactions remains an open question. Comparative analyses across platforms and sociopolitical contexts could clarify whether these dynamics are unique to YouTube or generalizable to other algorithm-driven spaces.

The relationship between online toxicity, self-censorship, and disengagement has profound implications for how we interpret ideological polarization and public discourse in digital spaces. Public discourse online may not accurately reflect individuals’ private beliefs or offline behaviors \cite{beauchamp2017predicting, nguyen2020association, schulz2020representing}. This discrepancy risks distorting our understanding of public opinion and intensifying polarization.

Nevertheless, as the boundaries between online and offline activities blur, especially in political discourse, understanding the role of social media platforms, recommendation algorithms, and collective antisocial behaviors becomes increasingly urgent. Social media platforms accelerate the dissemination of information, often reinforcing feedback loops that amplify toxicity and polarization. Our findings suggest a circular causality between toxicity and disengagement, where negative behaviors discourage meaningful participation, further entrenching polarization and alienation.

Addressing these dynamics is critical for preserving democratic engagement in a digital age. Platforms must consider how their design and algorithms influence discourse, fostering environments that support diverse perspectives and constructive dialogue. As online spaces increasingly shape public opinion and democratic processes, understanding and mitigating the effects of toxicity are essential steps toward ensuring their positive contribution to society.


\bmhead{Acknowledgments} We thank Ángel Sánchez, Felipe Vaca, and Onkar Sadekar for providing valuable feedback on earlier versions of the manuscript. G.J. and G.I. acknowledge support from CIVICA project ‘European Polarisation Observatory’ (EPO and EPO-WG). T.Y. acknowledges funding from Workday Limited. J.K. acknowledges funding from the Hungarian Academy of Sciences Lendület Program: LP2022-10/2022. G.I. acknowledges support from AFOSR (grant FA8655-20-1-7020), EU H2020 Humane AI-net (grant \#952026), and CHIST-ERA project SAI (grant FWF I 5205-N).

\bibliography{sn-article}


\begin{thebibliography}{66}
\ifx \bisbn   \undefined \def \bisbn  #1{ISBN #1}\fi
\ifx \binits  \undefined \def \binits#1{#1}\fi
\ifx \bauthor  \undefined \def \bauthor#1{#1}\fi
\ifx \batitle  \undefined \def \batitle#1{#1}\fi
\ifx \bjtitle  \undefined \def \bjtitle#1{#1}\fi
\ifx \bvolume  \undefined \def \bvolume#1{\textbf{#1}}\fi
\ifx \byear  \undefined \def \byear#1{#1}\fi
\ifx \bissue  \undefined \def \bissue#1{#1}\fi
\ifx \bfpage  \undefined \def \bfpage#1{#1}\fi
\ifx \blpage  \undefined \def \blpage #1{#1}\fi
\ifx \burl  \undefined \def \burl#1{\textsf{#1}}\fi
\ifx \doiurl  \undefined \def \doiurl#1{\url{https://doi.org/#1}}\fi
\ifx \betal  \undefined \def \betal{\textit{et al.}}\fi
\ifx \binstitute  \undefined \def \binstitute#1{#1}\fi
\ifx \binstitutionaled  \undefined \def \binstitutionaled#1{#1}\fi
\ifx \bctitle  \undefined \def \bctitle#1{#1}\fi
\ifx \beditor  \undefined \def \beditor#1{#1}\fi
\ifx \bpublisher  \undefined \def \bpublisher#1{#1}\fi
\ifx \bbtitle  \undefined \def \bbtitle#1{#1}\fi
\ifx \bedition  \undefined \def \bedition#1{#1}\fi
\ifx \bseriesno  \undefined \def \bseriesno#1{#1}\fi
\ifx \blocation  \undefined \def \blocation#1{#1}\fi
\ifx \bsertitle  \undefined \def \bsertitle#1{#1}\fi
\ifx \bsnm \undefined \def \bsnm#1{#1}\fi
\ifx \bsuffix \undefined \def \bsuffix#1{#1}\fi
\ifx \bparticle \undefined \def \bparticle#1{#1}\fi
\ifx \barticle \undefined \def \barticle#1{#1}\fi
\bibcommenthead
\ifx \bconfdate \undefined \def \bconfdate #1{#1}\fi
\ifx \botherref \undefined \def \botherref #1{#1}\fi
\ifx \url \undefined \def \url#1{\textsf{#1}}\fi
\ifx \bchapter \undefined \def \bchapter#1{#1}\fi
\ifx \bbook \undefined \def \bbook#1{#1}\fi
\ifx \bcomment \undefined \def \bcomment#1{#1}\fi
\ifx \oauthor \undefined \def \oauthor#1{#1}\fi
\ifx \citeauthoryear \undefined \def \citeauthoryear#1{#1}\fi
\ifx \endbibitem  \undefined \def \endbibitem {}\fi
\ifx \bconflocation  \undefined \def \bconflocation#1{#1}\fi
\ifx \arxivurl  \undefined \def \arxivurl#1{\textsf{#1}}\fi
\csname PreBibitemsHook\endcsname

\bibitem[\protect\citeauthoryear{Gonz{\'a}lez-Bail{\'o}n et~al.}{2023}]{gonzalez2023asymmetric}
\begin{barticle}
\bauthor{\bsnm{Gonz{\'a}lez-Bail{\'o}n}, \binits{S.}},
\bauthor{\bsnm{Lazer}, \binits{D.}},
\bauthor{\bsnm{Barber{\'a}}, \binits{P.}},
\bauthor{\bsnm{Zhang}, \binits{M.}},
\bauthor{\bsnm{Allcott}, \binits{H.}},
\bauthor{\bsnm{Brown}, \binits{T.}},
\bauthor{\bsnm{Crespo-Tenorio}, \binits{A.}},
\bauthor{\bsnm{Freelon}, \binits{D.}},
\bauthor{\bsnm{Gentzkow}, \binits{M.}},
\bauthor{\bsnm{Guess}, \binits{A.}}, \betal:
\batitle{Asymmetric ideological segregation in exposure to political news on facebook}.
\bjtitle{Science}
\bvolume{381}(\bissue{6656}),
\bfpage{392}--\blpage{398}
(\byear{2023})
\doiurl{10.1126/science.ade7138}
\end{barticle}
\endbibitem

\bibitem[\protect\citeauthoryear{Guess et~al.}{2023}]{guess2023social}
\begin{barticle}
\bauthor{\bsnm{Guess}, \binits{A.}},
\bauthor{\bsnm{Malhotra}, \binits{N.}},
\bauthor{\bsnm{Pan}, \binits{J.}},
\bauthor{\bsnm{Barber{\'a}}, \binits{P.}},
\bauthor{\bsnm{Allcott}, \binits{H.}},
\bauthor{\bsnm{Brown}, \binits{T.}},
\bauthor{\bsnm{Crespo-Tenorio}, \binits{A.}},
\bauthor{\bsnm{Dimmery}, \binits{D.}},
\bauthor{\bsnm{Freelon}, \binits{D.}},
\bauthor{\bsnm{Gentzkow}, \binits{M.}}, \betal:
\batitle{How do social media feed algorithms affect attitudes and behavior in an election campaign?}
\bjtitle{Science}
\bvolume{381}(\bissue{6656}),
\bfpage{398}--\blpage{404}
(\byear{2023})
\doiurl{10.1126/science.abp9364}
\end{barticle}
\endbibitem

\bibitem[\protect\citeauthoryear{Gimmler}{2001}]{gimmler2001deliberative}
\begin{barticle}
\bauthor{\bsnm{Gimmler}, \binits{A.}}:
\batitle{Deliberative democracy, the public sphere and the internet}.
\bjtitle{Philosophy \& Social Criticism}
\bvolume{27}(\bissue{4}),
\bfpage{21}--\blpage{39}
(\byear{2001})
\doiurl{10.1177/019145370102700402}
\end{barticle}
\endbibitem

\bibitem[\protect\citeauthoryear{Bressers and Hume}{2012}]{bressers2012mess}
\begin{barticle}
\bauthor{\bsnm{Bressers}, \binits{B.}},
\bauthor{\bsnm{Hume}, \binits{J.}}:
\batitle{Message boards, public discourse, and historical meaning: An online community reacts to {S}eptember 11}.
\bjtitle{American Journalism}
\bvolume{29}(\bissue{4}),
\bfpage{9}--\blpage{33}
(\byear{2012})
\doiurl{10.1080/08821127.2012.10677846}
\end{barticle}
\endbibitem

\bibitem[\protect\citeauthoryear{Lorenz-Spreen et~al.}{2023}]{lorenz2023systematic}
\begin{barticle}
\bauthor{\bsnm{Lorenz-Spreen}, \binits{P.}},
\bauthor{\bsnm{Oswald}, \binits{L.}},
\bauthor{\bsnm{Lewandowsky}, \binits{S.}},
\bauthor{\bsnm{Hertwig}, \binits{R.}}:
\batitle{A systematic review of worldwide causal and correlational evidence on digital media and democracy}.
\bjtitle{Nature human behaviour}
\bvolume{7}(\bissue{1}),
\bfpage{74}--\blpage{101}
(\byear{2023})
\doiurl{10.1038/s41562-022-01460-1}
\end{barticle}
\endbibitem

\bibitem[\protect\citeauthoryear{Asmelash}{26 July 2020}]{asmelash2020blm}
\begin{botherref}
\oauthor{\bsnm{Asmelash}, \binits{L.}}:
{How Black Lives Matter went from a hashtag to a global rallying cry}.
CNN
(26 July 2020).
\url{https://edition.cnn.com/2020/07/26/us/black-lives-matter-explainer-trnd/index.html}
\end{botherref}
\endbibitem

\bibitem[\protect\citeauthoryear{Frenkel}{1 January 2021}]{nyt2021capitol}
\begin{botherref}
\oauthor{\bsnm{Frenkel}, \binits{S.}}:
{The storming of Capitol Hill was organized on social media}.
New York Times
(1 January 2021).
\url{https://www.nytimes.com/2021/01/06/us/politics/protesters-storm-capitol-hill-building.html}
\end{botherref}
\endbibitem

\bibitem[\protect\citeauthoryear{Yasseri and Menczer}{2023}]{yasseri2023can}
\begin{barticle}
\bauthor{\bsnm{Yasseri}, \binits{T.}},
\bauthor{\bsnm{Menczer}, \binits{F.}}:
\batitle{Can crowdsourcing rescue the social marketplace of ideas?}
\bjtitle{Communications of the ACM}
\bvolume{66}(\bissue{9}),
\bfpage{42}--\blpage{45}
(\byear{2023})
\doiurl{10.1145/3578645}
\end{barticle}
\endbibitem

\bibitem[\protect\citeauthoryear{Stark et~al.}{26 May 2020}]{stark2020algorithms}
\begin{botherref}
\oauthor{\bsnm{Stark}, \binits{B.}},
\oauthor{\bsnm{Stegmann}, \binits{D.}},
\oauthor{\bsnm{Magin}, \binits{M.}},
\oauthor{\bsnm{J{\"u}rgens}, \binits{P.}}:
Are algorithms a threat to democracy? {T}he rise of intermediaries: {A} challenge for public discourse.
Algorithm Watch
(26 May 2020).
\url{https://algorithmwatch.org/de/wp-content/uploads/2020/05/Governing-Platforms-communications-study-Stark-May-2020-AlgorithmWatch.pdf}
\end{botherref}
\endbibitem

\bibitem[\protect\citeauthoryear{Bak-Coleman et~al.}{2021}]{bak2021stewardship}
\begin{barticle}
\bauthor{\bsnm{Bak-Coleman}, \binits{J.}},
\bauthor{\bsnm{Alfano}, \binits{M.}},
\bauthor{\bsnm{Barfuss}, \binits{W.}},
\bauthor{\bsnm{Bergstrom}, \binits{C.}},
\bauthor{\bsnm{Centeno}, \binits{M.}},
\bauthor{\bsnm{Couzin}, \binits{I.}},
\bauthor{\bsnm{Donges}, \binits{J.}},
\bauthor{\bsnm{Galesic}, \binits{M.}},
\bauthor{\bsnm{Gersick}, \binits{A.}},
\bauthor{\bsnm{Jacquet}, \binits{J.}}, \betal:
\batitle{Stewardship of global collective behavior}.
\bjtitle{Proceedings of the National Academy of Sciences}
\bvolume{118}(\bissue{27}),
\bfpage{2025764118}
(\byear{2021})
\doiurl{10.1073/pnas.2025764118}
\end{barticle}
\endbibitem

\bibitem[\protect\citeauthoryear{Porumbescu}{2017}]{porumbescu2017trust}
\begin{barticle}
\bauthor{\bsnm{Porumbescu}, \binits{G.}}:
\batitle{Not all bad news after all? {E}xploring the relationship between citizens’ use of online mass media for government information and trust in government}.
\bjtitle{International Public Management Journal}
\bvolume{20}(\bissue{3}),
\bfpage{409}--\blpage{441}
(\byear{2017})
\doiurl{10.1080/10967494.2016.1269859}
\end{barticle}
\endbibitem

\bibitem[\protect\citeauthoryear{Guess et~al.}{2021}]{guess2021partisan}
\begin{barticle}
\bauthor{\bsnm{Guess}, \binits{A.}},
\bauthor{\bsnm{Barber{\'a}}, \binits{P.}},
\bauthor{\bsnm{Munzert}, \binits{S.}},
\bauthor{\bsnm{Yang}, \binits{J.}}:
\batitle{The consequences of online partisan media}.
\bjtitle{Proceedings of the National Academy of Sciences}
\bvolume{118}(\bissue{14}),
\bfpage{2013464118}
(\byear{2021})
\doiurl{10.1073/pnas.2013464118}
\end{barticle}
\endbibitem

\bibitem[\protect\citeauthoryear{Schumann et~al.}{2022}]{schumann2022spiral}
\begin{barticle}
\bauthor{\bsnm{Schumann}, \binits{S.}},
\bauthor{\bsnm{Thomas}, \binits{F.}},
\bauthor{\bsnm{Ehrke}, \binits{F.}},
\bauthor{\bsnm{Bertlich}, \binits{T.}},
\bauthor{\bsnm{Dupont}, \binits{J.}}:
\batitle{Maintenance or change? {E}xamining the reinforcing spiral between social media news use and populist attitudes}.
\bjtitle{Information, Communication \& Society}
\bvolume{25}(\bissue{13}),
\bfpage{1934}--\blpage{1951}
(\byear{2022})
\doiurl{10.1080/1369118X.2021.1907435}
\end{barticle}
\endbibitem

\bibitem[\protect\citeauthoryear{{Casta\~o-Pulgar\'in} et~al.}{2021}]{castano2021hate}
\begin{barticle}
\bauthor{\bsnm{{Casta\~o-Pulgar\'in}}, \binits{S.}},
\bauthor{\bsnm{{Su\'arez-Betancur}}, \binits{N.}},
\bauthor{\bsnm{{Tilano-Vega}}, \binits{L.}},
\bauthor{\bsnm{{Herrera-L\'pez}}, \binits{H.}}:
\batitle{Internet, social media and online hate speech. {S}ystematic review}.
\bjtitle{Aggression and Violent Behavior}
\bvolume{58},
\bfpage{101608}
(\byear{2021})
\doiurl{10.1016/j.avb.2021.101608}
\end{barticle}
\endbibitem

\bibitem[\protect\citeauthoryear{Vidgen and Yasseri}{2020}]{vidgen2020detecting}
\begin{barticle}
\bauthor{\bsnm{Vidgen}, \binits{B.}},
\bauthor{\bsnm{Yasseri}, \binits{T.}}:
\batitle{Detecting weak and strong islamophobic hate speech on social media}.
\bjtitle{Journal of Information Technology \& Politics}
\bvolume{17}(\bissue{1}),
\bfpage{66}--\blpage{78}
(\byear{2020})
\doiurl{10.1080/19331681.2019.1702607}
\end{barticle}
\endbibitem

\bibitem[\protect\citeauthoryear{Bail et~al.}{2018}]{bail2018polarization}
\begin{barticle}
\bauthor{\bsnm{Bail}, \binits{C.}},
\bauthor{\bsnm{Argyle}, \binits{L.}},
\bauthor{\bsnm{Brown}, \binits{T.}},
\bauthor{\bsnm{Bumpus}, \binits{J.}},
\bauthor{\bsnm{Chen}, \binits{H.}},
\bauthor{\bsnm{Hunzaker}, \binits{F.}},
\bauthor{\bsnm{Lee}, \binits{J.}},
\bauthor{\bsnm{Mann}, \binits{M.}},
\bauthor{\bsnm{Merhout}, \binits{F.}},
\bauthor{\bsnm{Volfovsky}, \binits{A.}}:
\batitle{Exposure to opposing views on social media can increase political polarization}.
\bjtitle{Proceedings of the National Academy of Sciences}
\bvolume{115}(\bissue{37}),
\bfpage{9216}--\blpage{9221}
(\byear{2018})
\doiurl{10.1073/pnas.1804840115}
\end{barticle}
\endbibitem

\bibitem[\protect\citeauthoryear{Yarchi et~al.}{2021}]{yarchi2021digital}
\begin{barticle}
\bauthor{\bsnm{Yarchi}, \binits{M.}},
\bauthor{\bsnm{Baden}, \binits{C.}},
\bauthor{\bsnm{Kligler-Vilenchik}, \binits{N.}}:
\batitle{Political polarization on the digital sphere: {A} cross-platform, over-time analysis of interactional, positional, and affective polarization on social media}.
\bjtitle{Political Communication}
\bvolume{38}(\bissue{1-2}),
\bfpage{98}--\blpage{139}
(\byear{2021})
\doiurl{10.1080/10584609.2020.1785067}
\end{barticle}
\endbibitem

\bibitem[\protect\citeauthoryear{Bond et~al.}{2012}]{bond2012facebook}
\begin{barticle}
\bauthor{\bsnm{Bond}, \binits{R.}},
\bauthor{\bsnm{Fariss}, \binits{C.}},
\bauthor{\bsnm{Jones}, \binits{J.}},
\bauthor{\bsnm{Kramer}, \binits{A.}},
\bauthor{\bsnm{Marlow}, \binits{C.}},
\bauthor{\bsnm{Settle}, \binits{J.}},
\bauthor{\bsnm{Fowler}, \binits{J.}}:
\batitle{A 61-million-person experiment in social influence and political mobilization}.
\bjtitle{Nature}
\bvolume{489}(\bissue{7415}),
\bfpage{295}--\blpage{298}
(\byear{2012})
\doiurl{10.1038/nature11421}
\end{barticle}
\endbibitem

\bibitem[\protect\citeauthoryear{Boulianne}{2020}]{boulianne2020twenty}
\begin{barticle}
\bauthor{\bsnm{Boulianne}, \binits{S.}}:
\batitle{Twenty years of digital media effects on civic and political participation}.
\bjtitle{Communication Research}
\bvolume{47}(\bissue{7}),
\bfpage{947}--\blpage{966}
(\byear{2020})
\doiurl{10.1177/0093650218808186}
\end{barticle}
\endbibitem

\bibitem[\protect\citeauthoryear{Margetts et~al.}{2016}]{margetts2016political}
\begin{bbook}
\bauthor{\bsnm{Margetts}, \binits{H.}},
\bauthor{\bsnm{John}, \binits{P.}},
\bauthor{\bsnm{Hale}, \binits{S.}},
\bauthor{\bsnm{Yasseri}, \binits{T.}}:
\bbtitle{Political Turbulence: How Social Media Shape Collective Action}.
\bpublisher{Princeton University Press},
\blocation{Princeton}
(\byear{2016})
\end{bbook}
\endbibitem

\bibitem[\protect\citeauthoryear{Beaudoin}{2008}]{beaudoin2008internet}
\begin{barticle}
\bauthor{\bsnm{Beaudoin}, \binits{C.}}:
\batitle{The internet's impact on international knowledge}.
\bjtitle{New Media \& Society}
\bvolume{10}(\bissue{3}),
\bfpage{455}--\blpage{474}
(\byear{2008})
\doiurl{10.1177/1461444807085327}
\end{barticle}
\endbibitem

\bibitem[\protect\citeauthoryear{Park and Kaye}{2018}]{park2018engament}
\begin{barticle}
\bauthor{\bsnm{Park}, \binits{C.}},
\bauthor{\bsnm{Kaye}, \binits{B.}}:
\batitle{News engagement on social media and democratic citizenship: Direct and moderating roles of curatorial news use in political involvement}.
\bjtitle{Journalism \& Mass Communication Quarterly}
\bvolume{95}(\bissue{4}),
\bfpage{1103}--\blpage{1127}
(\byear{2018})
\doiurl{10.1177/1077699017753149}
\end{barticle}
\endbibitem

\bibitem[\protect\citeauthoryear{Skoric et~al.}{2016}]{skoric2016expression}
\begin{barticle}
\bauthor{\bsnm{Skoric}, \binits{M.}},
\bauthor{\bsnm{Zhu}, \binits{Q.}},
\bauthor{\bsnm{Pang}, \binits{N.}}:
\batitle{Social media, political expression, and participation in confucian asia}.
\bjtitle{Chinese Journal of Communication}
\bvolume{9}(\bissue{4}),
\bfpage{331}--\blpage{347}
(\byear{2016})
\doiurl{10.1080/17544750.2016.1143378}
\end{barticle}
\endbibitem

\bibitem[\protect\citeauthoryear{Dai et~al.}{2023}]{dai2023addressing}
\begin{bchapter}
\bauthor{\bsnm{Dai}, \binits{W.}},
\bauthor{\bsnm{Tao}, \binits{J.}},
\bauthor{\bsnm{Yan}, \binits{X.}},
\bauthor{\bsnm{Feng}, \binits{Z.}},
\bauthor{\bsnm{Chen}, \binits{J.}}:
\bctitle{Addressing unintended bias in toxicity detection: An lstm and attention-based approach}.
In: \bbtitle{2023 5th International Conference on Artificial Intelligence and Computer Applications (ICAICA)},
pp. \bfpage{375}--\blpage{379}
(\byear{2023}).
\doiurl{10.1109/ICAICA58456.2023.10405429} .
\bcomment{IEEE}
\end{bchapter}
\endbibitem

\bibitem[\protect\citeauthoryear{Oswald}{2023}]{oswald2023effects}
\begin{barticle}
\bauthor{\bsnm{Oswald}, \binits{L.}}:
\batitle{Effects of preemptive empathy interventions on reply toxicity among highly active social media users}.
\bjtitle{Public Discourse in Online Environments}
(\byear{2023})
\doiurl{10.31235/osf.io/evdxy}
\end{barticle}
\endbibitem

\bibitem[\protect\citeauthoryear{{Pew Research Center}}{December 2022}]{vogels2022teens}
\begin{botherref}
\oauthor{\bsnm{{Pew Research Center}}}:
Teens and cyberbullying 2022
(December 2022)
\end{botherref}
\endbibitem

\bibitem[\protect\citeauthoryear{{Pew Research Center}}{January 2021}]{vogels2021online}
\begin{botherref}
\oauthor{\bsnm{{Pew Research Center}}}:
The state of online harassment
(January 2021)
\end{botherref}
\endbibitem

\bibitem[\protect\citeauthoryear{Thomas et~al.}{2021}]{thomas2021hate}
\begin{bchapter}
\bauthor{\bsnm{Thomas}, \binits{K.}},
\bauthor{\bsnm{Akhawe}, \binits{D.}},
\bauthor{\bsnm{Bailey}, \binits{M.}},
\bauthor{\bsnm{Boneh}, \binits{D.}},
\bauthor{\bsnm{Bursztein}, \binits{E.}},
\bauthor{\bsnm{Consolvo}, \binits{S.}},
\bauthor{\bsnm{Dell}, \binits{N.}},
\bauthor{\bsnm{Durumeric}, \binits{Z.}},
\bauthor{\bsnm{Kelley}, \binits{P.}},
\bauthor{\bsnm{Kumar}, \binits{D.}}, \betal:
\bctitle{Sok: Hate, harassment, and the changing landscape of online abuse}.
In: \bbtitle{2021 IEEE Symposium on Security and Privacy (SP)},
pp. \bfpage{247}--\blpage{267}
(\byear{2021}).
\bcomment{IEEE}
\end{bchapter}
\endbibitem

\bibitem[\protect\citeauthoryear{Bail}{2022}]{bail2022breaking}
\begin{bbook}
\bauthor{\bsnm{Bail}, \binits{C.A.}}:
\bbtitle{Breaking the Social Media Prism: How to Make Our Platforms Less Polarizing}.
\bpublisher{Princeton University Press},
\blocation{Princeton}
(\byear{2022})
\end{bbook}
\endbibitem

\bibitem[\protect\citeauthoryear{Anderson et~al.}{2014}]{anderson2014nasty}
\begin{barticle}
\bauthor{\bsnm{Anderson}, \binits{A.A.}},
\bauthor{\bsnm{Brossard}, \binits{D.}},
\bauthor{\bsnm{Scheufele}, \binits{D.A.}},
\bauthor{\bsnm{Xenos}, \binits{M.A.}},
\bauthor{\bsnm{Ladwig}, \binits{P.}}:
\batitle{The “nasty effect:” online incivility and risk perceptions of emerging technologies}.
\bjtitle{Journal of computer-mediated communication}
\bvolume{19}(\bissue{3}),
\bfpage{373}--\blpage{387}
(\byear{2014})
\doiurl{10.1111/jcc4.12009}
\end{barticle}
\endbibitem

\bibitem[\protect\citeauthoryear{Kim et~al.}{2021}]{kim2021distorting}
\begin{barticle}
\bauthor{\bsnm{Kim}, \binits{J.}},
\bauthor{\bsnm{Guess}, \binits{A.}},
\bauthor{\bsnm{Nyhan}, \binits{B.}},
\bauthor{\bsnm{Reifler}, \binits{J.}}:
\batitle{The distorting prism of social media: How self-selection and exposure to incivility fuel online comment toxicity}.
\bjtitle{Journal of Communication}
\bvolume{71}(\bissue{6}),
\bfpage{922}--\blpage{946}
(\byear{2021})
\doiurl{10.1093/joc/jqab034}
\end{barticle}
\endbibitem

\bibitem[\protect\citeauthoryear{Bor and Petersen}{2022}]{bor2022psychology}
\begin{barticle}
\bauthor{\bsnm{Bor}, \binits{A.}},
\bauthor{\bsnm{Petersen}, \binits{M.B.}}:
\batitle{The psychology of online political hostility: A comprehensive, cross-national test of the mismatch hypothesis}.
\bjtitle{American political science review}
\bvolume{16}(\bissue{1}),
\bfpage{1}--\blpage{18}
(\byear{2022})
\doiurl{10.1017/S0003055421000885}
\end{barticle}
\endbibitem

\bibitem[\protect\citeauthoryear{Avalle et~al.}{2024}]{avalle2024persistent}
\begin{barticle}
\bauthor{\bsnm{Avalle}, \binits{M.}},
\bauthor{\bsnm{Di~Marco}, \binits{N.}},
\bauthor{\bsnm{Etta}, \binits{G.}},
\bauthor{\bsnm{Sangiorgio}, \binits{E.}},
\bauthor{\bsnm{Alipour}, \binits{S.}},
\bauthor{\bsnm{Bonetti}, \binits{A.}},
\bauthor{\bsnm{Alvisi}, \binits{L.}},
\bauthor{\bsnm{Scala}, \binits{A.}},
\bauthor{\bsnm{Baronchelli}, \binits{A.}},
\bauthor{\bsnm{Cinelli}, \binits{M.}}, \betal:
\batitle{Persistent interaction patterns across social media platforms and over time}.
\bjtitle{Nature}
\bvolume{628}(\bissue{8008}),
\bfpage{582}--\blpage{589}
(\byear{2024})
\doiurl{10.1038/s41586-024-07229-y}
\end{barticle}
\endbibitem

\bibitem[\protect\citeauthoryear{Wulczyn et~al.}{2017}]{wulczyn2017ex}
\begin{bchapter}
\bauthor{\bsnm{Wulczyn}, \binits{E.}},
\bauthor{\bsnm{Thain}, \binits{N.}},
\bauthor{\bsnm{Dixon}, \binits{L.}}:
\bctitle{Ex machina: Personal attacks seen at scale}.
In: \bbtitle{Proceedings of the 26th International Conference on World Wide Web},
pp. \bfpage{1391}--\blpage{1399}
(\byear{2017}).
\doiurl{10.1145/3038912.3052591}
\end{bchapter}
\endbibitem

\bibitem[\protect\citeauthoryear{Lees et~al.}{2022}]{lees2022new}
\begin{bchapter}
\bauthor{\bsnm{Lees}, \binits{A.}},
\bauthor{\bsnm{Tran}, \binits{V.}},
\bauthor{\bsnm{Tay}, \binits{Y.}},
\bauthor{\bsnm{Sorensen}, \binits{J.}},
\bauthor{\bsnm{Gupta}, \binits{J.}},
\bauthor{\bsnm{Metzler}, \binits{D.}},
\bauthor{\bsnm{Vasserman}, \binits{L.}}:
\bctitle{A new generation of perspective api: Efficient multilingual character-level transformers}.
In: \bbtitle{Proceedings of the 28th ACM SIGKDD Conference on Knowledge Discovery and Data Mining},
pp. \bfpage{3197}--\blpage{3207}
(\byear{2022}).
\doiurl{10.1145/3534678.3539147}
\end{bchapter}
\endbibitem

\bibitem[\protect\citeauthoryear{}{}]{perspective2024how}
\begin{botherref}
Perspective API - How it works.
\url{https://www.perspectiveapi.com/how-it-works/}.
Accessed: 21 October 2024
\end{botherref}
\endbibitem

\bibitem[\protect\citeauthoryear{Vidgen and Derczynski}{2020}]{vidgen2020directions}
\begin{barticle}
\bauthor{\bsnm{Vidgen}, \binits{B.}},
\bauthor{\bsnm{Derczynski}, \binits{L.}}:
\batitle{Directions in abusive language training data, a systematic review: Garbage in, garbage out}.
\bjtitle{Plos one}
\bvolume{15}(\bissue{12}),
\bfpage{0243300}
(\byear{2020})
\doiurl{10.1371/journal.pone.0243300}
\end{barticle}
\endbibitem

\bibitem[\protect\citeauthoryear{Jigsaw}{9 March 2018}]{medium2018bias}
\begin{botherref}
\oauthor{\bsnm{Jigsaw}}:
{Unintended Bias and Identity Terms}.
Medium
(9 March 2018).
\url{https://medium.com/jigsaw/unintended-bias-and-names-of-frequently-targeted-groups-8e0b81f80a23}
\end{botherref}
\endbibitem

\bibitem[\protect\citeauthoryear{Nogara et~al.}{2023}]{nogara2023toxic}
\begin{botherref}
\oauthor{\bsnm{Nogara}, \binits{G.}},
\oauthor{\bsnm{Pierri}, \binits{F.}},
\oauthor{\bsnm{Cresci}, \binits{S.}},
\oauthor{\bsnm{Luceri}, \binits{L.}},
\oauthor{\bsnm{T{\"o}rnberg}, \binits{P.}},
\oauthor{\bsnm{Giordano}, \binits{S.}}:
Toxic Bias: Perspective API misreads German as more toxic.
Preprint at \url{https://arxiv.org/abs/2312.12651}
(2023)
\end{botherref}
\endbibitem

\bibitem[\protect\citeauthoryear{Ramage}{2007}]{ramage2007hmm}
\begin{botherref}
\oauthor{\bsnm{Ramage}, \binits{D.}}:
CS229: Hidden Markov Models Fundamentals.
Standford University.
\url{https://cs229.stanford.edu/section/cs229-hmm.pdf}
(2007)
\end{botherref}
\endbibitem

\bibitem[\protect\citeauthoryear{Eisner}{2002}]{eisner2002interactive}
\begin{bchapter}
\bauthor{\bsnm{Eisner}, \binits{J.}}:
\bctitle{An interactive spreadsheet for teaching the forward-backward algorithm}.
In: \bbtitle{Proceedings of the ACL-02 Workshop on Effective Tools and Methodologies for Teaching Natural Language Processing and Computational Linguistics},
pp. \bfpage{10}--\blpage{18}
(\byear{2002}).
\bcomment{\url{https://aclanthology.org/W02-0102.pdf}}
\end{bchapter}
\endbibitem

\bibitem[\protect\citeauthoryear{Istrail}{2020}]{istrail2020hmm}
\begin{botherref}
\oauthor{\bsnm{Istrail}, \binits{S.}}:
HMM: The Learning Problem.
Department of Computer Science, Brown University.
\url{https://cs.brown.edu/courses/csci1820/resources/HMMs_The_Learning_Problem_slides.pdf}
(2020)
\end{botherref}
\endbibitem

\bibitem[\protect\citeauthoryear{Hosseinmardi et~al.}{2021}]{hosseinmardi2021examining}
\begin{barticle}
\bauthor{\bsnm{Hosseinmardi}, \binits{H.}},
\bauthor{\bsnm{Ghasemian}, \binits{A.}},
\bauthor{\bsnm{Clauset}, \binits{A.}},
\bauthor{\bsnm{Mobius}, \binits{M.}},
\bauthor{\bsnm{Rothschild}, \binits{D.}},
\bauthor{\bsnm{Watts}, \binits{D.}}:
\batitle{Examining the consumption of radical content on youtube}.
\bjtitle{Proceedings of the National Academy of Sciences}
\bvolume{118}(\bissue{32}),
\bfpage{2101967118}
(\byear{2021})
\doiurl{10.1073/pnas.2101967118}
\end{barticle}
\endbibitem

\bibitem[\protect\citeauthoryear{Haroon et~al.}{2023}]{haroon2023auditing}
\begin{barticle}
\bauthor{\bsnm{Haroon}, \binits{M.}},
\bauthor{\bsnm{Wojcieszak}, \binits{M.}},
\bauthor{\bsnm{Chhabra}, \binits{A.}},
\bauthor{\bsnm{Liu}, \binits{X.}},
\bauthor{\bsnm{Mohapatra}, \binits{P.}},
\bauthor{\bsnm{Shafiq}, \binits{Z.}}:
\batitle{Auditing youtube’s recommendation system for ideologically congenial, extreme, and problematic recommendations}.
\bjtitle{Proceedings of the National Academy of Sciences}
\bvolume{120}(\bissue{50}),
\bfpage{2213020120}
(\byear{2023})
\doiurl{10.1073/pnas.221302012}
\end{barticle}
\endbibitem

\bibitem[\protect\citeauthoryear{Mastrine}{1 February 2019}]{mastrine2019bias}
\begin{botherref}
\oauthor{\bsnm{Mastrine}, \binits{J.}}:
{Introducing the AllSides Media Bias Chart}.
AllSides
(1 February 2019).
\url{https://www.allsides.com/blog/introducing-allsides-media-bias-chart}
\end{botherref}
\endbibitem

\bibitem[\protect\citeauthoryear{}{}]{perspective2024attributes}
\begin{botherref}
Perspective API - Attributes \& Languages.
\url{https://developers.perspectiveapi.com/s/about-the-api-attributes-and-languages?language=en_US}.
Accessed: 21 October 2024
\end{botherref}
\endbibitem

\bibitem[\protect\citeauthoryear{}{}]{perspective2024score}
\begin{botherref}
Perspective API - Score.
\url{https://developers.perspectiveapi.com/s/about-the-api-score?language=en_US}.
Accessed: 21 October 2024
\end{botherref}
\endbibitem

\bibitem[\protect\citeauthoryear{Saveski et~al.}{2021}]{saveski2021structure}
\begin{bchapter}
\bauthor{\bsnm{Saveski}, \binits{M.}},
\bauthor{\bsnm{Roy}, \binits{B.}},
\bauthor{\bsnm{Roy}, \binits{D.}}:
\bctitle{The structure of toxic conversations on twitter}.
In: \beditor{\bsnm{{Leskovec, J., \textit{et al}}}} (ed.)
\bbtitle{Proceedings of the Web Conference 2021},
pp. \bfpage{1086}--\blpage{1097}.
\bpublisher{Association for Computing Machinery},
\blocation{New YorkNYUnited States}
(\byear{2021}).
\doiurl{10.1145/3442381.344986}
\end{bchapter}
\endbibitem

\bibitem[\protect\citeauthoryear{Sheth et~al.}{2022}]{sheth2022defining}
\begin{barticle}
\bauthor{\bsnm{Sheth}, \binits{A.}},
\bauthor{\bsnm{Shalin}, \binits{V.L.}},
\bauthor{\bsnm{Kursuncu}, \binits{U.}}:
\batitle{Defining and detecting toxicity on social media: context and knowledge are key}.
\bjtitle{Neurocomputing}
\bvolume{490},
\bfpage{312}--\blpage{318}
(\byear{2022})
\doiurl{10.1016/j.neucom.2021.11.095}
\end{barticle}
\endbibitem

\bibitem[\protect\citeauthoryear{Rajadesingan et~al.}{2020}]{rajadesingan2020quick}
\begin{bchapter}
\bauthor{\bsnm{Rajadesingan}, \binits{A.}},
\bauthor{\bsnm{Resnick}, \binits{P.}},
\bauthor{\bsnm{Budak}, \binits{C.}}:
\bctitle{Quick, community-specific learning: How distinctive toxicity norms are maintained in political subreddits}.
In: \bbtitle{Proceedings of the International AAAI Conference on Web and Social Media},
pp. \bfpage{557}--\blpage{568}
(\byear{2020}).
\doiurl{10.1609/icwsm.v14i1.7323}
\end{bchapter}
\endbibitem

\bibitem[\protect\citeauthoryear{Hopp et~al.}{2020}]{hopp2020correlating}
\begin{barticle}
\bauthor{\bsnm{Hopp}, \binits{T.}},
\bauthor{\bsnm{Vargo}, \binits{C.J.}},
\bauthor{\bsnm{Dixon}, \binits{L.}},
\bauthor{\bsnm{Thain}, \binits{N.}}:
\batitle{Correlating self-report and trace data measures of incivility: A proof of concept}.
\bjtitle{Social Science Computer Review}
\bvolume{38}(\bissue{5}),
\bfpage{584}--\blpage{599}
(\byear{2020})
\doiurl{10.1177/089443931881424}
\end{barticle}
\endbibitem

\bibitem[\protect\citeauthoryear{{AllSides Staff}}{3 January 2023}]{staff2023score}
\begin{botherref}
\oauthor{\bsnm{{AllSides Staff}}}:
{Introducing the AllSides Media Bias Meter™}.
AllSides
(3 January 2023).
\url{https://www.allsides.com/blog/introducing-allsides-media-bias-meter}
\end{botherref}
\endbibitem

\bibitem[\protect\citeauthoryear{Brady et~al.}{2017}]{brady2017emotion}
\begin{barticle}
\bauthor{\bsnm{Brady}, \binits{W.}},
\bauthor{\bsnm{Wills}, \binits{J.}},
\bauthor{\bsnm{Jost}, \binits{J.}},
\bauthor{\bsnm{Tucker}, \binits{J.}},
\bauthor{\bsnm{Van~Bavel}, \binits{J.}}:
\batitle{Emotion shapes the diffusion of moralized content in social networks}.
\bjtitle{Proceedings of the National Academy of Sciences}
\bvolume{114}(\bissue{28}),
\bfpage{7313}--\blpage{7318}
(\byear{2017})
\doiurl{10.1073/pnas.1618923114}
\end{barticle}
\endbibitem

\bibitem[\protect\citeauthoryear{McLaughlin}{28 September 2020}]{cnn2020btaylor}
\begin{botherref}
\oauthor{\bsnm{McLaughlin}, \binits{E.}}:
{Anger erupts in American cities after charging decision in Breonna Taylor case}.
CNN
(28 September 2020).
\url{https://edition.cnn.com/2020/09/28/us/weekend-protests-breonna-taylor/index.html}
\end{botherref}
\endbibitem

\bibitem[\protect\citeauthoryear{{BBC Staff}}{22 April 2021}]{bbc2021blm}
\begin{botherref}
\oauthor{\bsnm{{BBC Staff}}}:
{George Floyd: Timeline of black deaths and protests}.
BBC
(22 April 2021).
\url{https://www.bbc.com/news/world-us-canada-52905408}
\end{botherref}
\endbibitem

\bibitem[\protect\citeauthoryear{Huang}{22 February 2021}]{huang2021covid}
\begin{botherref}
\oauthor{\bsnm{Huang}, \binits{P.}}:
{'A Loss To The Whole Society': U.S. COVID-19 Death Toll Reaches 500,000}.
NPR
(22 February 2021).
\url{https://www.npr.org/sections/health-shots/2021/02/22/969494791/a-loss-to-the-whole-society-u-s-covid-19-death-toll-reaches-500-000}
\end{botherref}
\endbibitem

\bibitem[\protect\citeauthoryear{Pohle et~al.}{2017}]{pohle2017selecting}
\begin{barticle}
\bauthor{\bsnm{Pohle}, \binits{J.}},
\bauthor{\bsnm{Langrock}, \binits{R.}},
\bauthor{\bsnm{Van~Beest}, \binits{F.M.}},
\bauthor{\bsnm{Schmidt}, \binits{N.M.}}:
\batitle{Selecting the number of states in hidden markov models: pragmatic solutions illustrated using animal movement}.
\bjtitle{Journal of Agricultural, Biological and Environmental Statistics}
\bvolume{22},
\bfpage{270}--\blpage{293}
(\byear{2017})
\doiurl{10.1007/s13253-017-0283-8}
\end{barticle}
\endbibitem

\bibitem[\protect\citeauthoryear{Vayansky and Kumar}{2020}]{vayansky2020review}
\begin{barticle}
\bauthor{\bsnm{Vayansky}, \binits{I.}},
\bauthor{\bsnm{Kumar}, \binits{S.}}:
\batitle{A review of topic modeling methods}.
\bjtitle{Information Systems}
\bvolume{94},
\bfpage{101582}
(\byear{2020})
\doiurl{10.1016/j.is.2020.101582}
\end{barticle}
\endbibitem

\bibitem[\protect\citeauthoryear{Gerlach et~al.}{2018}]{gerlach2018network}
\begin{barticle}
\bauthor{\bsnm{Gerlach}, \binits{M.}},
\bauthor{\bsnm{Peixoto}, \binits{T.P.}},
\bauthor{\bsnm{Altmann}, \binits{E.G.}}:
\batitle{A network approach to topic models}.
\bjtitle{Science Advances}
\bvolume{4}(\bissue{7}),
\bfpage{1360}
(\byear{2018})
\doiurl{10.1126/sciadv.aaq1360}
\end{barticle}
\endbibitem

\bibitem[\protect\citeauthoryear{Blex and Yasseri}{2022}]{blex2022positive}
\begin{barticle}
\bauthor{\bsnm{Blex}, \binits{C.}},
\bauthor{\bsnm{Yasseri}, \binits{T.}}:
\batitle{Positive algorithmic bias cannot stop fragmentation in homophilic networks}.
\bjtitle{The Journal of Mathematical Sociology}
\bvolume{46}(\bissue{1}),
\bfpage{80}--\blpage{97}
(\byear{2022})
\doiurl{10.1080/0022250X.2020.1818078}
\end{barticle}
\endbibitem

\bibitem[\protect\citeauthoryear{Falkenberg et~al.}{2024}]{falkenberg2024patterns}
\begin{barticle}
\bauthor{\bsnm{Falkenberg}, \binits{M.}},
\bauthor{\bsnm{Zollo}, \binits{F.}},
\bauthor{\bsnm{Quattrociocchi}, \binits{W.}},
\bauthor{\bsnm{Pfeffer}, \binits{J.}},
\bauthor{\bsnm{Baronchelli}, \binits{A.}}:
\batitle{Patterns of partisan toxicity and engagement reveal the common structure of online political communication across countries}.
\bjtitle{Nature Communications}
\bvolume{15}(\bissue{1}),
\bfpage{9560}
(\byear{2024})
\doiurl{10.1038/s41467-024-53868-0}
\end{barticle}
\endbibitem

\bibitem[\protect\citeauthoryear{Beauchamp}{2017}]{beauchamp2017predicting}
\begin{barticle}
\bauthor{\bsnm{Beauchamp}, \binits{N.}}:
\batitle{Predicting and interpolating state-level polls using twitter textual data}.
\bjtitle{American Journal of Political Science}
\bvolume{61}(\bissue{2}),
\bfpage{490}--\blpage{503}
(\byear{2017})
\doiurl{10.1111/ajps.12274}
\end{barticle}
\endbibitem

\bibitem[\protect\citeauthoryear{Nguyen et~al.}{2020}]{nguyen2020association}
\begin{barticle}
\bauthor{\bsnm{Nguyen}, \binits{T.T.}},
\bauthor{\bsnm{Adams}, \binits{N.}},
\bauthor{\bsnm{Huang}, \binits{D.}},
\bauthor{\bsnm{Glymour}, \binits{M.M.}},
\bauthor{\bsnm{Allen}, \binits{A.M.}},
\bauthor{\bsnm{Nguyen}, \binits{Q.C.}}:
\batitle{The association between state-level racial attitudes assessed from twitter data and adverse birth outcomes: Observational study}.
\bjtitle{JMIR public health and surveillance}
\bvolume{6}(\bissue{3}),
\bfpage{17103}
(\byear{2020})
\doiurl{10.2196/17103}
\end{barticle}
\endbibitem

\bibitem[\protect\citeauthoryear{Schulz et~al.}{2020}]{schulz2020representing}
\begin{botherref}
\oauthor{\bsnm{Schulz}, \binits{W.S.}},
\oauthor{\bsnm{Guess}, \binits{A.M.}},
\oauthor{\bsnm{Barber{\'a}}, \binits{P.}},
\oauthor{\bsnm{Munzert}, \binits{S.}},
\oauthor{\bsnm{Gottlieb}, \binits{A.}},
\oauthor{\bsnm{Hughes}, \binits{A.}},
\oauthor{\bsnm{Remy}, \binits{E.}},
\oauthor{\bsnm{Shah}, \binits{S.}},
\oauthor{\bsnm{Smith}, \binits{A.}}:
{(Mis)representing} ideology on twitter: How social influence shapes online political expression
(2020).
\url{https://simonmunzert.github.io/meof/ material/schulz-et-al-ideology-twitter-apsa.pdf}
\end{botherref}
\endbibitem

\bibitem[\protect\citeauthoryear{Jo et~al.}{2012}]{jo2012circadian}
\begin{barticle}
\bauthor{\bsnm{Jo}, \binits{H.}},
\bauthor{\bsnm{Karsai}, \binits{M.}},
\bauthor{\bsnm{Kert{\'e}sz}, \binits{J.}},
\bauthor{\bsnm{Kaski}, \binits{K.}}:
\batitle{Circadian pattern and burstiness in mobile phone communication}.
\bjtitle{New Journal of Physics}
\bvolume{14}(\bissue{1}),
\bfpage{013055}
(\byear{2012})
\doiurl{10.1088/1367-2630/14/1/013055}
\end{barticle}
\endbibitem

\bibitem[\protect\citeauthoryear{Folador et~al.}{2019}]{folador2019protein}
\begin{bchapter}
\bauthor{\bsnm{Folador}, \binits{E.}},
\bauthor{\bsnm{Tiwari}, \binits{S.}},
\bauthor{\bsnm{{Da Paz Barbosa}}, \binits{C.}},
\bauthor{\bsnm{Jamal}, \binits{S.}},
\bauthor{\bsnm{{Da Costa Schulze}}, \binits{M.}},
\bauthor{\bsnm{Barh}, \binits{D.}},
\bauthor{\bsnm{Azevedo}, \binits{V.}}:
\bctitle{Protein-protein interactions: an overview}.
In: \beditor{\bsnm{Ranganathan}, \binits{S.}},
\beditor{\bsnm{Gribskov}, \binits{M.}},
\beditor{\bsnm{Nakai}, \binits{K.}},
\beditor{\bsnm{Schönbach}, \binits{C.}} (eds.)
\bbtitle{Encyclopedia of Bioinformatics and Computational Biology},
pp. \bfpage{821}--\blpage{833}.
\bpublisher{Academic Press},
\blocation{Oxford}
(\byear{2019}).
\doiurl{10.1016/B978-0-12-809633-8.20292-6}
\end{bchapter}
\endbibitem

\end{thebibliography}

\newpage
\begin{center}
{\LARGE Supplementary Information for}\\[0.7cm]
{\Large \textbf{Toxic behavior silences online political conversations}}\\[0.5cm]
{\large G. Juncosa*, T. Yasseri, J. Koltia, G. Iñiguez*}\\[0.7cm]
{\small $^*$Corresponding authors email: juncosa\_maria@phd.ceu.edu, gerardo.iniguez@tuni.fi}\\[2cm]

\end{center}

\section*{S1. Data Description}\label{app:data-description}
\begin{table}[h!]
\caption{\textbf{Descriptive Statistics by Comment Type.} The dataset contains top-level comments and replies from videos posted by six prominent U.S. news outlets chosen for their active YouTube comment sections and for representing a broad spectrum of political ideologies. The time frame begins in September 2020, and extends through April 2021.}\label{app:desciptive-stats}
\begin{tabular*}{\textwidth}{@{\extracolsep\fill}llcccccccc@{}}
\toprule
                           &                             & \multicolumn{8}{c}{Comments}                                                                                                  \\ \midrule
                           & \multicolumn{1}{c}{Channel} & Top Level & $\%$\footnotemark[1] & Toxicity\footnotemark[2] & Insult\footnotemark[3] & Replies  & $\%$ & Toxicity & Insult \\ \midrule
\multirow{3}{*}{Lean left} & ABC News                    & 1993690   & 49.8                                    & 20.3     & 20.8   & 2009172  & 50.2                 & 17.1     & 18.1   \\
                           & CBS                         & 1443344   & 49.1                                    & 6.5      & 6.8    & 1496646  & 50.9                 & 4.8      & 5.4    \\
                           & CNN                         & 5945976   & 59.0                                    & 12.0     & 11.0   & 4128018  & 41.0                 & 7.6      & 7.3    \\ \midrule
\multirow{3}{*}{Right}     & OAN                         & 403054    & 65.8                                    & 6.7      & 6.0    & 209545   & 34.2                 & 5.5      & 5.7    \\
                           & Newsmax                     & 1881808   & 66.4                                    & 7.0      & 7.9    & 950237   & 33.6                 & 7.0      & 8.2    \\
                           & Fox News                    & 7387795   & 62.9                                    & 10.0     & 10.5   & 4355601  & 37.1                 & 8.0      & 8.9    \\
                           & ALL                         & 19055667  & 59.2                                    & 11.1     & 11.1   & 13149219 & 40.8                 & 8.8      & 9.3    \\ \bottomrule
\end{tabular*}
\footnotetext[1]{As a percentage of all comments in channel}
\footnotetext[2]{Percentage of toxic comments by comment type. A comment is toxic if its toxicity score is greater than 0.6.}
\footnotetext[3]{Percentage of insulting comments by comment type. A comment is insulting if its insult score is greater than 0.6.}
\end{table}

\begin{figure}[H]
\centering
\includegraphics[width=0.6\textwidth]{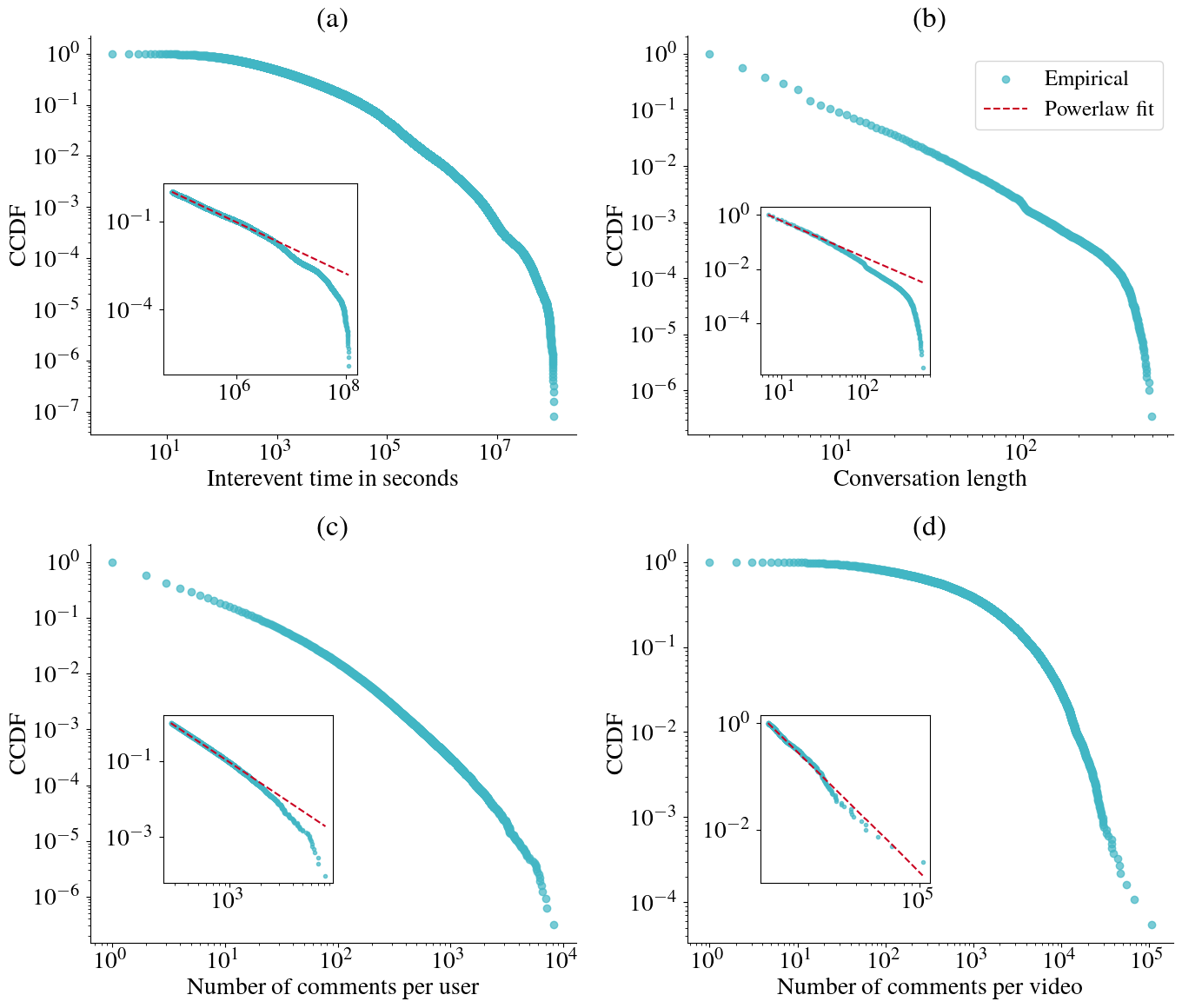}
\caption{\textbf{Complementary Cumulative Distribution Functions (CCDF) for engagement metrics in YouTube}. Figures show empirical data (blue dots) and power law fits (red dashed lines) for each plot. Insets highlight alignment with power law distributions at different scales. \textbf{(a)} Distribution of interevent times (in seconds) for user commenting behavior. The distribution is heterogeneous, as evidenced by the heavy-tailed CCDF. The presence of both short and long interevent times suggests burstiness in commenting frequency, rather than a consistent interaction rate. \textbf{(b)} Distribution of conversation length, measured by the number of comments per conversation. This distribution follows a heavy-tailed pattern, indicating that while most conversations are brief, a few are lengthy and maintain high engagement levels. \textbf{(c)} Distribution of the number of comments per user, illustrating user activity levels. The CCDF shows a heterogeneous and heavy-tailed distribution, with a few users being highly active while the majority post occasionally. \textbf{(d)} CCDF of the number of comments per video, illustrating overall engagement levels per video. This distribution underscores the nature of content virality in online platforms, where specific videos become popular and attract a significant number of comments.}\label{app: ccdf_engagementMetrics}
\end{figure}

Figure \ref{app: ccdf_engagementMetrics} presents distribution functions for key metrics in the YouTube dataset. Panel (a) shows the distribution of interevent times, measured in seconds, for commenting behavior. Reflecting typical human communication patterns, which are often bursty and marked by irregular interaction intervals \cite{jo2012circadian}, our dataset shows both short and long interevent times, indicating a bursty commenting frequency rather than a steady interaction rate. Panel (b) shows the distribution of conversation length, measured by the number of comments per conversation. Generally, conversations have between 2 and 500 comments, with only a few sustaining high engagement, while the majority remain relatively brief. Panel (c) shows the number of comments per user, an indicator of user activity level, while panel (d) shows the number of comments per video, reflecting overall engagement levels. Both metrics reveal similar heterogeneous patterns: a small number of users are highly active, while most post only occasionally, and specific videos gain popularity, attracting a large volume of comments.

\section*{S2. Bias Ratings}\label{app:bias-ratings}

\begin{table}[!htbp]
\caption{\textbf{AllSides Media Bias Ratings.} The AllSides Media Bias Meter provides a nuanced rating of media outlets on a scale from -6 (farthest Left) to +6 (farthest Right). The bias meter rating is based on a combination of methods, including Editorial Reviews conducted by a multipartisan panel and a Blind Bias Survey where respondents rate outlets on an 11-point scale. The final rating typically averages results from these methods, with adjustments made for data consistency and recency, and more recent reviews carrying greater weight.}\label{tab2}
\begin{tabular*}{\textwidth}{@{\extracolsep\fill}cccc@{}}
\toprule
\multicolumn{1}{c}{Channel} & Bias Meter Rating & Bias Meter Score & Rating Confidence \\ \midrule
ABC\footnotemark[1]                         & Lean Left         & -2,40            & High              \\
CBS\footnotemark[2]                         & Lean Left         & -1,50            & High              \\
CNN\footnotemark[3]                         & Lean Left         & -1,30            & High              \\
OAN\footnotemark[4]                         & Right             & 3,10             & Medium            \\
Newsmax\footnotemark[5]                     & Right             & 3,28             & Low/Initial       \\
Fox News\footnotemark[6]                    & Right             & 3,88             & Medium            \\ \bottomrule
\end{tabular*}
\footnotetext[1]{\url{https://www.allsides.com/news-source/abc-news-media-bias} Retrieved August 22, 2024}
\footnotetext[2]{\url{https://www.allsides.com/news-source/cbs-news-media-bias} Retrieved August 22, 2024}
\footnotetext[3]{\url{https://www.allsides.com/news-source/cnn-media-bias} Retrieved August 22, 2024}
\footnotetext[4]{\url{https://www.allsides.com/news-source/one-america-news-network-media-bias} Retrieved August 22, 2024}
\footnotetext[5]{\url{https://www.allsides.com/news-source/newsmax} Retrieved August 22, 2024}
\footnotetext[6]{\url{https://www.allsides.com/news-source/fox-news-media-bias} Retrieved August 22, 2024}
\end{table}

In January 2023, AllSides launched the AllSides Media Bias Meter. Before this update, the website classified media outlets into five bias categories: Left, Lean Left, Center, Lean Right, and Right. The new bias meter provides a more detailed assessment, rating outlets on a scale from -6 to +6, with 0.0 representing a perfect Center, -6 indicating the farthest Left, and +6 indicating the farthest Right.

The bias meter rating is derived from a combination of methods. First, AllSides conducts Editorial Reviews with a multipartisan panel representing different political views. Each panelist reviews news content individually, identifies bias indicators, and provides their initial rating. The panel then discusses their findings, and members may revise their ratings. The ratings are averaged by grouping them into three categories (Left, Center, Right) and calculating a weighted average. If all panelists agree that the final rating should differ from the calculated average, they can unanimously override it.

In addition to Editorial Reviews, AllSides conducts a Blind Bias Survey to assess bias meter ratings. Respondents from different political perspectives rate headlines and news reports on an 11-point Likert scale. AllSides then calculates an average score for each bias group and an overall weighted average to produce the Blind Bias Survey result. This result uses a scale of -9 to +9, which differs from the AllSides Media Bias Meter scale of -6 to +6. Despite the possibility of higher scores in the survey, AllSides caps the final bias rating at -6 and +6 to align with the methodology of other analyses like Editorial Reviews.

The Final Bias Meter rating is determined by the number and timing of review methods applied to a source. If multiple reviews have been conducted, AllSides considers all data and assigns an overall numerical value, usually averaging results from Editorial Reviews and Blind Bias Surveys. Adjustments may be made based on factors like data consistency and recency, with full documentation of any changes. If only one methodology is available, the rating will reflect that result. When multiple methods were applied at different times, more recent reviews are given greater weight in calculating the final rating.

\section*{S3. Perspective API prevalence all attributes}\label{app:perspective-prevalence}

\begin{figure}[H]
\centering
\includegraphics[scale=0.45]{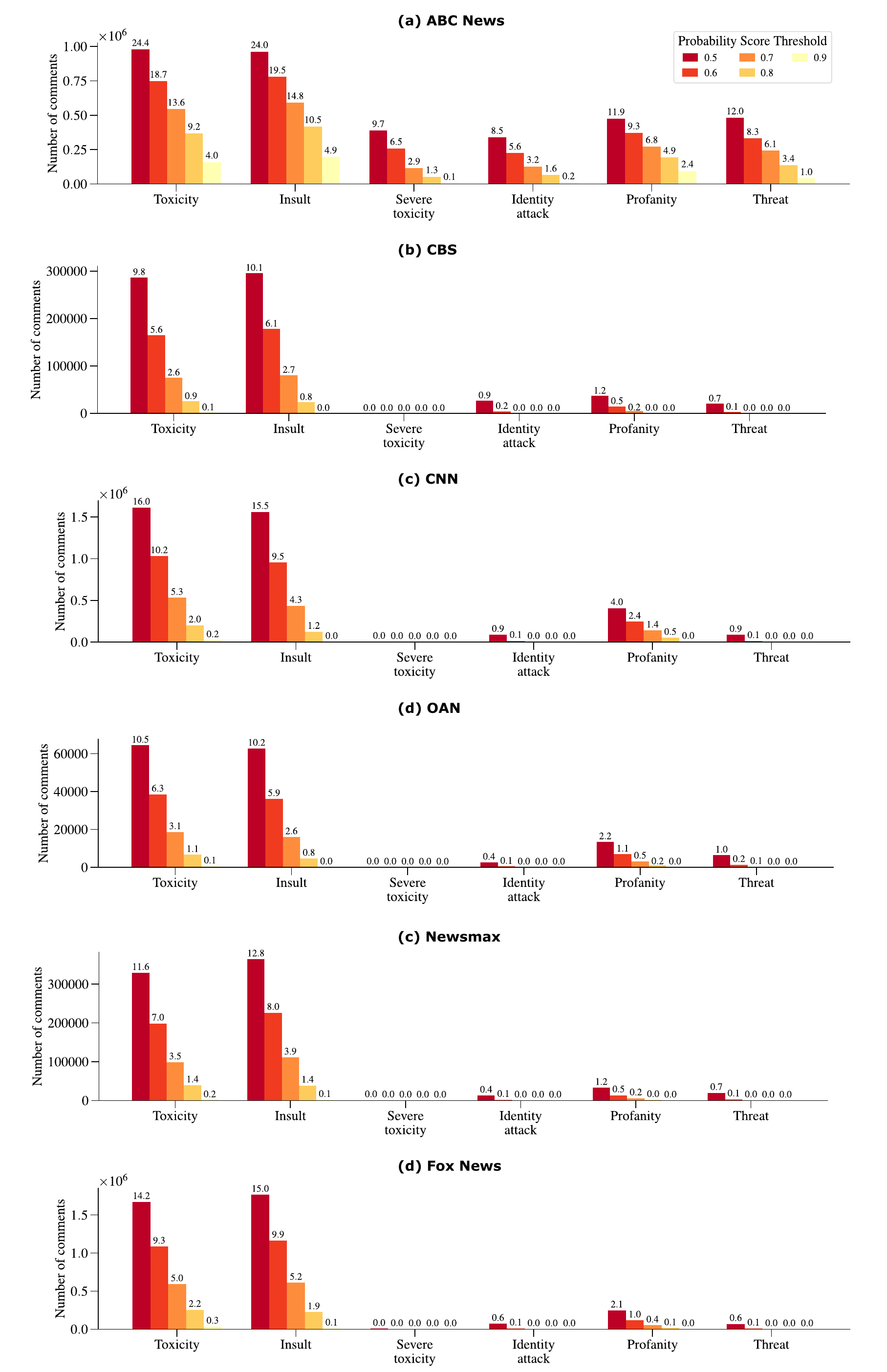}
\caption{Prevalence of six emotions by channel in the dataset. The height of the bars indicates the absolute frequency of each emotion, with percentages shown at the top. Bar colors represent different probability score thresholds.}\label{sentPrevalence}
\end{figure}

\section*{S4. Hidden Markov Model fit details}\label{app:hmm-Nstates}

\begin{table}[!ht]
\caption{Number of Hidden  Markov model fits}\label{app:hmm-n-realizations}
\begin{tabular}{@{}llcccc@{}}
\toprule
\multicolumn{1}{c}{\multirow{2}{*}{}} & \multicolumn{1}{c}{\multirow{2}{*}{Channel}} & \multicolumn{2}{c}{Threads} & \multicolumn{2}{c}{No. Fits} \\ \cmidrule(l){3-6} 
\multicolumn{1}{c}{}   & \multicolumn{1}{c}{} & N       & $\%$                       & Toxicity & Insult \\ \midrule
Lean left              & ABC News             & 447756  & \multicolumn{1}{c}{15.5}  & 4139     & 3993   \\
                       & CBS                  & 356538  & \multicolumn{1}{c}{12.4}  & 4179     & 3443   \\
                       & CNN                  & 888010  & \multicolumn{1}{c}{30.8}  & 4758     & 4391   \\ \midrule
\multirow{3}{*}{Right} & OAN                  & 48318   & \multicolumn{1}{c}{1.7}   & 4680     & 4703   \\
                       & Newsmax              & 187440  & \multicolumn{1}{c}{6.5}   & 3698     & 3817   \\
                       & Fox News             & 955409  & \multicolumn{1}{c}{33.1}  & 3410     & 3248   \\ \midrule
                       & ALL                  & 2883471 & \multicolumn{1}{c}{100.0} &          &        \\ \bottomrule
\end{tabular}
\end{table}

A Hidden Markov Model (HMM) is a statistical framework designed to analyze systems where the observed data is generated by underlying states that are not directly visible. As a specific form of a dynamic Bayesian network \cite{folador2019protein}, HMMs excel at modeling temporal and sequential data. First introduced in speech recognition, HMMs have since been successfully applied to various fields, including biological sequence analysis, handwriting recognition, and protein-protein interaction predictions \cite{folador2019protein}. These models operate on the principle of the Markov process, where the future state depends solely on the present state, not the sequence of past states. However, unlike traditional Markov models, HMMs introduce hidden states, meaning the actual states of the system cannot be observed directly. Instead, only observable symbols or events generated by these states are accessible \cite{ramage2007hmm,istrail2020hmm}.

An HMM consists of several essential components. Hidden states form the backbone of the model, representing the unobservable conditions influencing the observed data. Observable data, such as a sequence of symbols or events, serves as the manifestation of these hidden states. The model incorporates transition probabilities, which define the likelihood of moving from one hidden state to another, and emission probabilities, which specify the likelihood of observing particular data given a specific hidden state. Together, these parameters enable the HMM to model the relationship between observable data and the hidden processes generating it \cite{folador2019protein, ramage2007hmm, istrail2020hmm}.

Training an HMM involves estimating these probabilities to maximize the likelihood of observed sequences, typically using iterative algorithms such as Expectation-Maximization. Once trained, the model is capable of performing several critical tasks. It can evaluate sequences to determine how well they align with the training data, decode observed sequences to identify the most likely sequence of hidden states, and even generate new sequences that mirror the patterns of the training data. Despite its efficiency in solving these problems, finding the "best" set of probabilities is computationally challenging, as it involves solving an NP-hard optimization problem \cite{folador2019protein}.

HMMs are particularly useful for applications requiring probabilistic modeling of unobservable dynamics. For instance, in speech recognition, HMMs map sound patterns to spoken words, while in biological research, they predict gene sequences or protein interactions based on observed biological data \cite{folador2019protein, ramage2007hmm, istrail2020hmm}. A conceptual example of their use involves inferring weather conditions from observed activities, such as beach visits or staying indoors. In this case, while the actual weather remains hidden, the HMM can deduce the most probable weather patterns based on the observed behaviors \cite{eisner2002interactive}.

This work leverages the utility of Hidden Markov Models (HMMs) as a robust analytical framework for making inferences about systems with hidden dynamics. By effectively bridging the gap between observable data and unobservable processes, HMMs serve as a powerful tool for understanding and predicting patterns in sequential data. In this study, we apply HMMs to explore the relationship between negative sentiment and disengagement in conversations occurring under YouTube videos.

In this work, we employ HMMs to \textit{learn} model parameters in various contexts. Specifically, we infer the transition and emission probabilities that best align with the sequences of comments, or conversations, in our dataset, tailoring the parameters for each of the six channels. Our approach focuses on learning these parameters for groups of videos, allowing us to capture and analyze aggregated behavior across video collections. We organize videos in two distinct ways: (1) by grouping them according to their news media channel publishers, and (2) by categorizing them based on the topics they address, as identified through topic modeling, regardless of their channel of origin. We use two alternative methods for defining video ensembles to ensure that the observed relationship between negative sentiment and disengagement is not dependent on the specific group definitions.

To achieve this, we fitted a two-level HMM, as illustrated in Figure \ref{HMMDiagram}. In this model, ${X}_{1}=0$ represents the absence of activity or the conclusion of a conversation, ${X}_{2}=1$ corresponds to a non-toxic or non-insulting post, and ${X}_{3}=2$ signifies a comment categorized as either toxic or insulting. To convert our conversation threads into a sequence of observations, we began by classifying each comment as either toxic or insulting (${X}{3}=2$) or non-toxic/non-insulting (${X}{2}=1$). This process transforms a chronologically ordered text sequence into a sequence of 1s and 2s. Additionally, we appended a zero (${X}_{1}=0$) to the end of every sequence/conversation to indicate its conclusion, ensuring that all conversations end with a 0.

Due to the size of our dataset, a sampling process was necessary, since fitting the model parameters for most ensembles required too many conversations to be processed at once. Through experimentation, we determined that no more than 45,000 conversations could be used at a time for training. On average, fitting the model parameters for one realization required approximately two hours. To address this, we sampled 45,000 conversations with replacement for each realization, splitting each sample into a training set (80\%, or 36,000 conversations) and a test set (20\%, or 9,000 conversations). We conducted as many realizations as our computational resources allowed, and Table \ref{app:hmm-n-realizations} provides a summary of the number of model fits included in our results.

\begin{figure}[h]
\caption{Negative log-likelihood for component candidates, for conversations on ABC news}\label{app:log-likelihood-abc}
\includegraphics[scale=0.90]{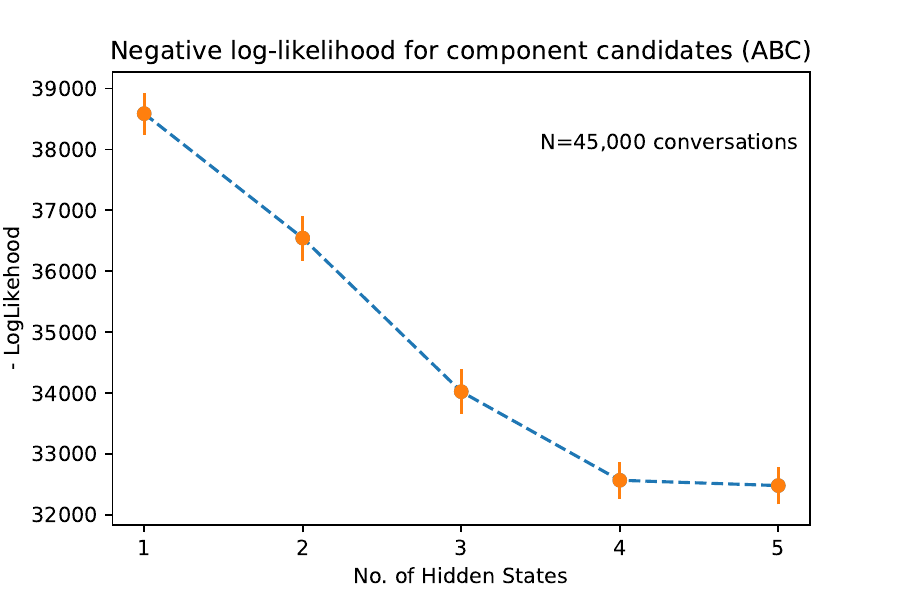}
\centering
\end{figure}

While we fixed the number of hidden states at two, we conducted robustness checks with models using three to five states. These checks showed that the four-state model performed best, as indicated by its log-likelihood (see Figure \ref{app:log-likelihood-abc}). However, A qualitative comparison of the two-state and four-state models revealed that the four-state model primarily provides a more detailed characterization of state (${Z}{2}$), which represents the tone of the conversation preceding disengagement, where disengagement is not possible (i.e., $P({X}{1}|{Z}_{2}=0)$). While this refinement is informative, the primary aim of this study is to distinguish between self-censorship and active engagement. A detailed exploration of the variations within state ${Z}_{2}$ falls outside the scope of this work. Furthermore, our preference for the simpler two-state model aligns with prior research showing that model selection procedures for HMMs often overfit by favoring models with an excessive number of states \cite{pohle2017selecting}.

\end{document}